\documentclass{aa}  

\usepackage{graphicx}
\usepackage{txfonts}
\usepackage{natbib}
\bibpunct{(}{)}{;}{a}{}{,}

\begin{document}

\title{Modelling the photosphere of active stars for planet detection and characterization}

\author{Enrique Herrero\inst{\ref{inst1}} \and Ignasi Ribas\inst{\ref{inst1}}
\and Carme Jordi\inst{\ref{inst2}} \and Juan Carlos Morales\inst{\ref{inst3}}
\and Manuel Perger\inst{\ref{inst1}} \and Albert Rosich\inst{\ref{inst1},\ref{inst2},4}
} 
\institute{Institut de Ci\`{e}ncies de l'Espai (CSIC-IEEC), Campus UAB, Carrer
de Can Magrans s/n, 08193 Cerdanyola del Vall\`es, Spain,
\email{eherrero@ice.cat, iribas@ice.cat, perger@ice.cat,
rosich@ice.cat}\label{inst1} 
\and
Dept. d'Astronomia i Meteorologia, Institut de Ci\`{e}ncies del Cosmos (ICC),
Universitat de Barcelona (IEEC-UB), Mart\'{i} Franqu\`{e}s 1, E08028 Barcelona, Spain, 
\email{carme.jordi@ub.edu}\label{inst2}
\and
LESIA-Observatoire de Paris, CNRS, UPMC Univ. Paris 06, Univ. Paris-Diderot, 5 Pl. Jules Janssen, 92195 Meudon CEDEX, France,
\email{Juan-Carlos.Morales@obspm.fr}\label{inst3}
\and
Reial Acadèmia de Ciències i Arts de Barcelona (RACAB),  Barcelona, 
Spain}\label{inst4}

\date{Received <date> /
Accepted <date>}

\abstract
{Stellar activity patterns are responsible for jitter effects that are 
observed at different timescales and amplitudes in the measurements 
obtained from photometric and spectroscopic time series observations. 
 These effects are currently in the focus of many exoplanet search
projects, since the lack of a well-defined characterization and correction
strategy hampers the detection of the signals associated with small
exoplanets.}
{Accurate simulations of the stellar photosphere based on the most recent
available models for main sequence stars can provide synthetic photometric and
spectroscopic time series data. These may help to investigate the relation
between activity jitter and stellar parameters when considering different
active region patterns. Moreover, jitters can be analysed at different 
wavelength scales (defined by the passbands of given instruments or 
space missions) in order to design strategies to remove or minimize 
them.}
{In this work we present the \texttt{StarSim} tool, which is based on a model
for a spotted rotating photosphere built from the integration of the spectral
contribution of a fine grid of surface elements. The model includes all
significant effects affecting the flux intensities and the wavelength of
spectral features produced by active regions and planets. The
resulting synthetic time series data generated with this simulator are used in
order to characterize the effects of activity jitter in extrasolar planet
measurements from photometric and spectroscopic observations.}
{Several cases of synthetic data series for Sun-like stars are presented to
illustrate the capabilities of the methodology. A specific application for the
characterization and modelling of the spectral signature of active regions is
considered, showing that the chromatic effects of faculae are dominant for low
temperature contrasts of spots. Synthetic multi-band photometry and radial
velocity time series are modelled for HD~189733 by adopting the known system
parameters and fitting for the map of active regions with StarSim. Our
algorithm reproduces both the photometry and the RVs to good precision,
generally better than the studies published to date. We evaluate the RV
signature of the activity in HD~189733 by exploring a grid of solutions from
the photometry. We find that the use of RV data in the inverse problem could
break degeneracies and allow for a better determination of some stellar and
activity parameters, e.g., the configuration of active regions, the temperature
contrast of spots and the amount of faculae. In addition, the effects of spots
are studied for a set of simulated transit photometry, showing that these can
introduce variations in $R_{p}/R_{*}$ measurements with a spectral signature
and amplitude which are very similar to the signal of an atmosphere dominated
by dust.

}{}

\keywords{stars: activity --
  stars: rotation -- stars: starspots} 
\maketitle

\section{Introduction}

Stellar activity in late-type main sequence stars induces photometric
modulations and apparent radial velocity (thereafter RV) variations that
may hamper the detection of Earth-like planets \citep{Lagrange10,Meunier10} and
the measurement of their transit parameters \citep{Barros13}, mass and
atmospheric properties.  A number of recent studies have been focused on
trying to separate the small exoplanetary signals and the variations in the
stellar signal both on light and RV curves
\citep{Dumusque14b,Haywood14,Robertson14}. However, we still lack detailed
understanding of these stellar variations from the astrophysical point of
view.

Late-type stars (i.e., late-F, G, K and M spectral types) are known to be
variable to some extent due to the effects of activity. These effects are seen
in the stellar photosphere in the form of spots and faculae, but we lack an accurate 
model of their signature on time series photometric and spectroscopic data.
The signal of these activity patterns is modulated by the stellar rotation
period. However, other phenomena also exist that may induce periodic signals in
photometric and RV data. Pulsations, granulation, long term evolution of active
regions and magnetic cycles may limit the capabilities of some planet searches.
Such effects need to be subtracted or accounted for the design of survey
strategies \citep{Dumusque11a,Dumusque11b,Moulds13}.

Significant improvement in our knowledge of activity effects on 
starlight will be crucial to make the most out of present and future planet search
spectrographs (HARPS, HARPS-N, CARMENES, ESPRESSO, HIRES, APF, SPIRou) and transit
observations from space (CHEOPS, PLATO, JWST). The most up-to-date and
comprehensive information on stellar variability comes from studies of
the Kepler mission, which are based on the observation and analysis of
$\sim$150,000 stars taken from the first Kepler data release. \cite{Ciardi11}
have found that 80$\%$ of M dwarfs have light dispersion less than 500 ppm over
a period of 12 hours, while G dwarfs are the most stable group down to 40 ppm.
 \cite{McQuillan12} investigates the variability properties of main
sequence stars in the Kepler data, finding that the typical amplitude and
time-scale increase towards later spectral types, which could be related to an
increase in the characteristic size and life-time of active regions. Kepler
operates in the visible (430 to 890 nm) where stellar photometric variability
is at least a factor of 2 higher than in the near and thermal IR (the ``sweet 
spot'' for the characterization of exoplanet atmospheres) due to the
increasing contrast between spots and the stellar photosphere with 
decreasing wavelength. Timescales for stellar activity are generally 
different from those associated with single transit observations (a few 
hours) and so removal of this spectral variability is possible. As a 
case in point, photometric modulations in the host of CoRoT-7 b are of 
the order of 2$\%$ \citep{Lanza10} and yet a transit with a depth of 
0.03$\%$ was identified \citep{Leger09}. Analysis of observations from 
Kepler have yielded comparable results. 


The impact of stellar activity effects is very different in the case of primary
(transit) and occultation observations.  Alterations in the spot distribution
across the stellar surface can modify the transit depth because of
possible spot crossing events and also the changing ratio of photosphere and
spotted areas on the face of the star \citep{Oshagh13b,Daassou14}. This can
give rise to spurious planetary radius variations when multiple transit
observations are considered. Correction of this effect requires the use of very
quiet stars or precise modelling of the stellar surface using external
constraints \citep{McCullough14,Oshagh13,Oshagh14}. The situation is much
simpler for occultations, where the planetary emission follows directly from
the transit depth measurement. In this case, only activity-induced variations on the
timescale of the duration of the occultation need to be corrected to ensure
that the proper stellar flux baseline is used. In the particular case of
exoplanet characterization space missions, photometric monitoring in the
visible will help in the correction of activity effects in the near and thermal
IR, where the planet signal is higher.

The effects of stellar activity on time series data have also been the subject of several recent studies using different methods for modelling
observations of spotted stars
\citep{Lanza03,Lanza10,Desort07,Aigrain12,Meunier10,Kipping12}. Examples are the SOAP
\citep{Boisse12} and SOAP 2.0 \citep{Dumusque14a} codes, which implement
surface integration techniques to reproduce the cross-correlation function
(CCF) of an active star by using a Gaussian function or real observed
solar CCFs. These include the effects of Doppler shifts and convective
blueshift inhibition produced by spots and faculae, as well as the limb
brightening effect of faculae and a quadratic limb darkening law for the 
quiet photosphere. As discussed by \cite{Dumusque14a}, it is essential to 
reproduce convective blueshift effects from real line shapes to obtain accurate
RV variations. However, most of the codes only consider the
effect caused by the flux differential of active regions
\citep{Boisse12,Oshagh13,Oshagh14}. \cite{Desort07} use Kurucz models together with a black-body approximation to reproduce the flux effect of spots on photometry and radial velocities, but they do not model the effects of convection inhibition in active regions. Moreover, most works reproduce the spectroscopic measurements from a single spectral line or
generate the CCF from a simple gaussian model. These types of models do
not preserve all the spectral information of spots and faculae, and therefore
do not allow for the study of RV jitter produced at different spectral ranges,
and the chromatic flux variations induced by activity.

In this work we present the \texttt{StarSim} tool, which investigates
 the effects of stellar activity on spectroscopic or
spectrophotometric observations by simulating full
spectra from the spotted photosphere of a rotating star. We 
use atmosphere models for low mass stars to generate synthetic spectra 
for the stellar surface, including the quiet photosphere, spots and faculae. The spectrum of the entire 
visible face of the star is obtained by summing the contribution of a 
grid of small surface elements and by considering their individual 
signals. Using such simulator, time series spectra can be obtained 
covering any time interval (e.g., the rotation period of 
the star or longer). By multiplying the spectra
 with the specific transfer function, both accounted for the atmospherical and the instrumental response, the methodology can be used to 
study the chromatic effects of spots and faculae on photometric 
modulations and spectroscopic jitters. The results are used to 
investigate methodologies to correct or mitigate the effects of activity 
on spectrophotometric time series data. Some of these methods are presented and discussed considering the case of HD 189733 in Sect.\ref{hd_ex}. Our results will allow us to 
evaluate the effects of activity patterns on the stellar 
flux and hence define the best strategies to optimize exoplanet searches 
and measurement experiments.

This paper is the first of a series presenting several activity simulating and
modelling applications of \texttt{StarSim} to improve our capabilities for planet
detection and characterization.

\section{Simulation of the photosphere of active stars}
\label{starsim}
\subsection{The models}
\label{starsimmodels}

\subsubsection{Phoenix synthetic spectra}
\label{phoenix}

\begin{figure}[]
\begin{center}
\includegraphics[width=\columnwidth]{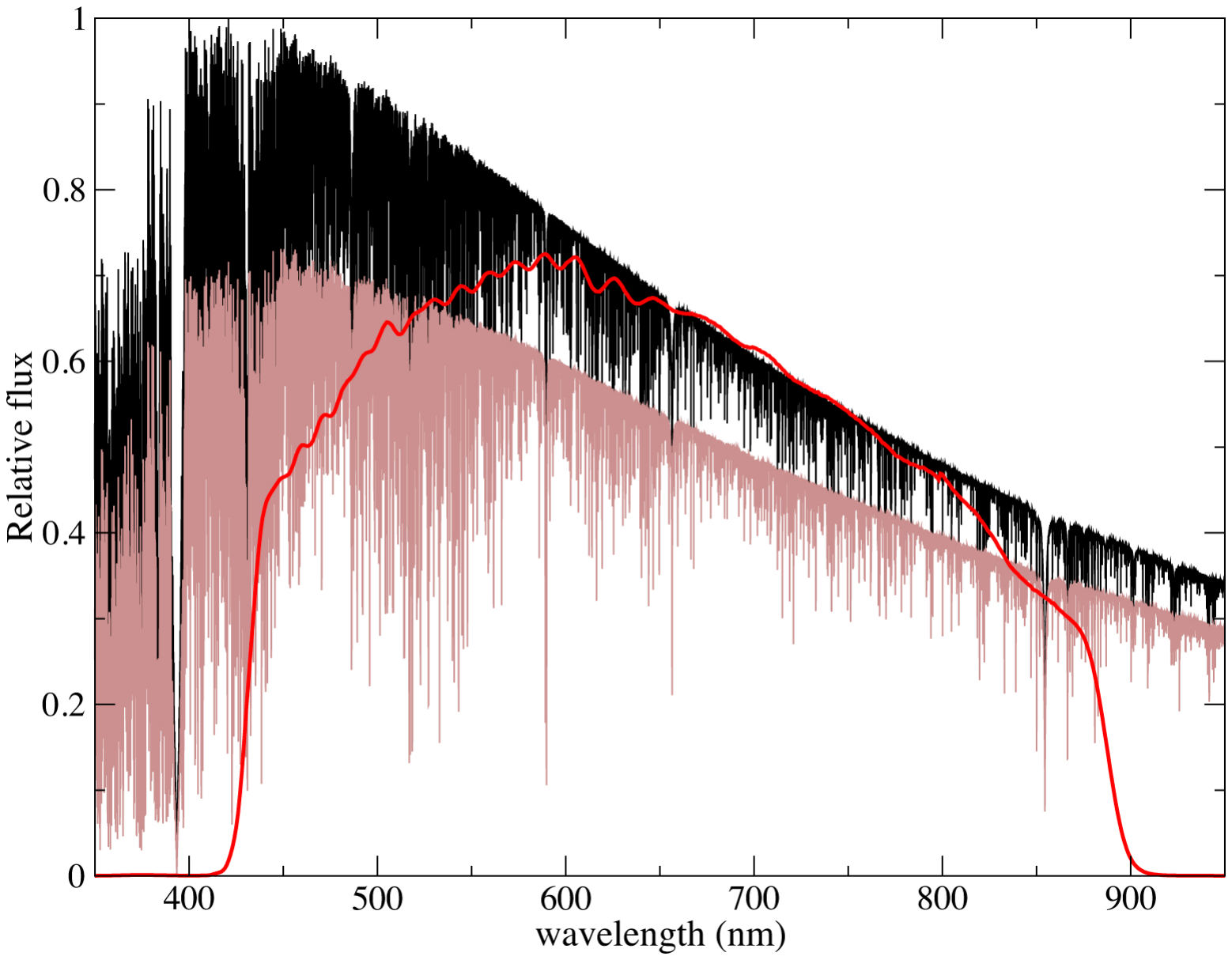}
\end{center}
\caption{Solar-like star synthetic spectrum for $T_{\rm eff}=$5770 K and $\log 
g=4.5$ (black) generated with BT-Settl models, compared to a synthetic spectrum
with a $T_{\rm eff}$ value 310~K lower (brown), which could be representative
of a spot. In red, the transmission function of the Kepler passband.
}
\label{fig_sun}
\end{figure}

Among the most recent model atmosphere grids, we use spectra from the 
BT-Settl database \citep{Allard13} generated with the Phoenix code. These
are used to reproduce the spectral signal for the different elements in the 
photosphere (quiet photosphere, spots and faculae) by considering
different temperatures for the synthetic spectra. Models include revised solar
oxygen abundances \citep{Caffau11} and a cloud formation recipe, which
manage to reproduce the photometric and spectroscopic properties also for very
low mass stars. In our simulations, LTE models are used both for the quiet
photosphere and the active regions (spots and faculae).  NLTE models from
\cite{Fontenla09} will be considered in future versions of the code to model
faculae, as they have shown to better reproduce the spectral irradiance of
activity features over all the magnetic cycle in the case of the Sun
\citep{Fontenla11}. Synthetic spectra from BT-Settl models are available for
2600~K $<T_{\rm eff}<$ 70000~K, +3.5 $\le \log g \le $ +5.0 and several values
of alpha enhancement and metallicity. An example for a solar-type star is shown 
in Fig.~\ref{fig_sun}.

These models include the effects of convection as described by the 
mixing length theory of turbulent transport \citep{Vitense53} 
characterized by the mixing length parameter $\alpha$ as discussed by 
\cite{Ludwig99} using 3D radiative models. Convection has a significant 
impact on line profiles, which are modified when active regions cross 
the stellar surface, and hence it is responsible for a significant part 
of the jitter observed in RV measurements 
\citep{Meunier10}.

A BT-Settl database of spectra is currently available for resolutions 
defined by a sampling $\ge$0.05{\AA}, which is not enough to study line 
profiles and obtain accurate RV measurements. In order to 
perform simulations for RV jitter studies and line 
profiles, high resolution spectra presented by \cite{Husser13} are used 
instead. The library is also based on the Phoenix code, and the 
synthetic spectra cover the wavelength range from 500 to 5500
nm with resolutions of $R=500,000$ in the optical and near IR, 
$R=100,000$ in the IR and $\Delta\lambda=0.1\AA$ in the UV. The 
parameter space covers 2300~K $\le T_{\rm eff} \le $ 12000~K, 0.0 $\le 
\log g \le $ +6.0, -4.0 $\le [$Fe/H$] \le $ +1.0, and 0.2 $\le 
[\alpha/$Fe$] \le $ +1.2.

\subsubsection{Solar spectra}
\label{mod_sol}

\begin{figure}[]
\begin{center}
\includegraphics[width=\columnwidth]{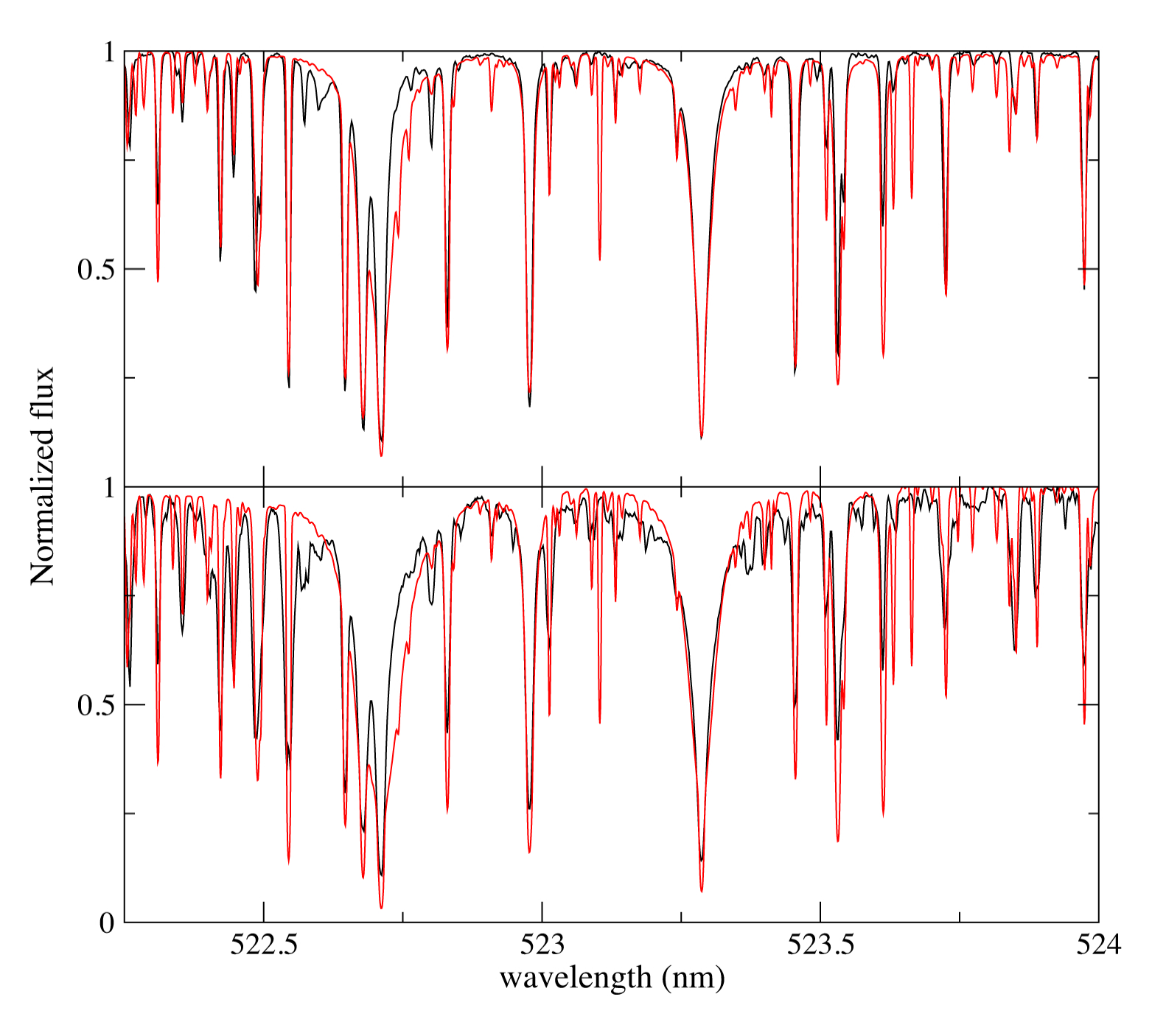}
\end{center}
\caption{Top: Comparison between an observed high-resolution spectrum 
for the solar photosphere (black) and a Phoenix synthetic spectrum (red) from
\cite{Husser13} using $T_{\rm eff}=$5770~K, $\log g=4.5$ and solar abundances.
Bottom: Same for the observed spectrum of a spot region (black), now compared
with a Phoenix synthetic spectrum of $T_{\rm eff}=$5460~K.
}
\label{comp_sun}
\end{figure}

In addition to synthetic spectra from model atmospheres, observed 
high-resolution spectra of the Sun are used in order to test the results 
coming from the Phoenix models. The spectra are obtained from a photospheric 
region and a spot with the 1-meter Fourier Transform Spectrometer of the 
National Solar Observatory located at Kitt Peak, and have a resolution 
of $R\sim10^6$ covering wavelengths from 390 nm to 665 nm. Figure~\ref{comp_sun} 
shows a comparison between the solar observed spectra and synthetic 
Phoenix spectra from \cite{Husser13} for a Sun-like star, both for the 
quiet photosphere and for a spot region. The temperature of the photospheric 
region was fixed to $T_{\rm eff}=$5770~K for the synthetic spectrum, which
 agrees with most of the features in the observed data.  In the case
of the spot spectrum, $T_{\rm eff}$ was varied from 4500 to 5500~K, finding the
best agreement for 5460~K (temperature contrast $\Delta T_{\rm spot}$=310~K)
when considering the whole wavelength range. This was done by
minimizing the rms of the different models divided by the spectrum of the
spot. The result is consistent with the measurements made by \cite{Eker03}, who determined $\Delta T_{\rm spot}\simeq$300~K for the combination of umbra and penumbra by analysing multi-channel fluxes of an equatorial passage of a sunspot.

\subsection{Data simulation}

\subsubsection{Model and stellar parameters}
\label{params}

The stellar photosphere is divided into a grid of surface elements whose 
physical and geometrical properties are described individually. The space 
resolution of the grid is an input parameter of the program, and 
$1^{\circ}\times1^{\circ}$ size elements have been found to be adequate 
to correctly reproduce most active region configurations and transiting 
planet effects down to $\sim10^{-6}$ photometric precision. A complete list of
the input parameters of the program is presented in Table~\ref{tab:simparams}.
 The table also shows the typical values considered for each parameter,
based on the grid coverage of the models and on some physical constraints that
are discussed below.

The \texttt{StarSim} tool is able to reproduce any possible distribution of active
regions that can be defined with the adopted surface grid. Active regions with
complex geometric properties can be defined by adding the contribution of
small circular ones. The total
number of active regions, as well as their surface and time distributions, can
be provided as an input by the user or randomly generated by the program
following different statistics defined in the input parameters. The number of
regions at a given time depends on a flat distribution between a specified
minimum, $Nr_{\rm min}$, and a maximum, $Nr_{\rm max}$. The sizes of the spots
follow a Gaussian distribution, defined by a mean area $\bar{A}_{\rm Sn}$ and a
standard deviation $\sigma(A_{\rm Sn})$, in units of the total stellar surface.
Therefore, the mean filling factor ($\bar{FF}$) over time in the whole stellar
surface is determined by the mean number of spots, $\bar{Nr}=(Nr_{\rm
max}+Nr_{\rm min})/2$ and their mean area by:

\begin{equation}
\bar{FF}=\bar{Nr} \cdot \bar{A}_{\rm Sn}
\label{meanff}
\end{equation}

Note that this filling factor only varies due to active region 
evolution and is different from the projected filling factor, which also 
accounts for the geometric projection of spots due to stellar 
inclination and rotation modulation.

\begin{figure}[]
\begin{center}
\includegraphics[width=\columnwidth,angle=0]{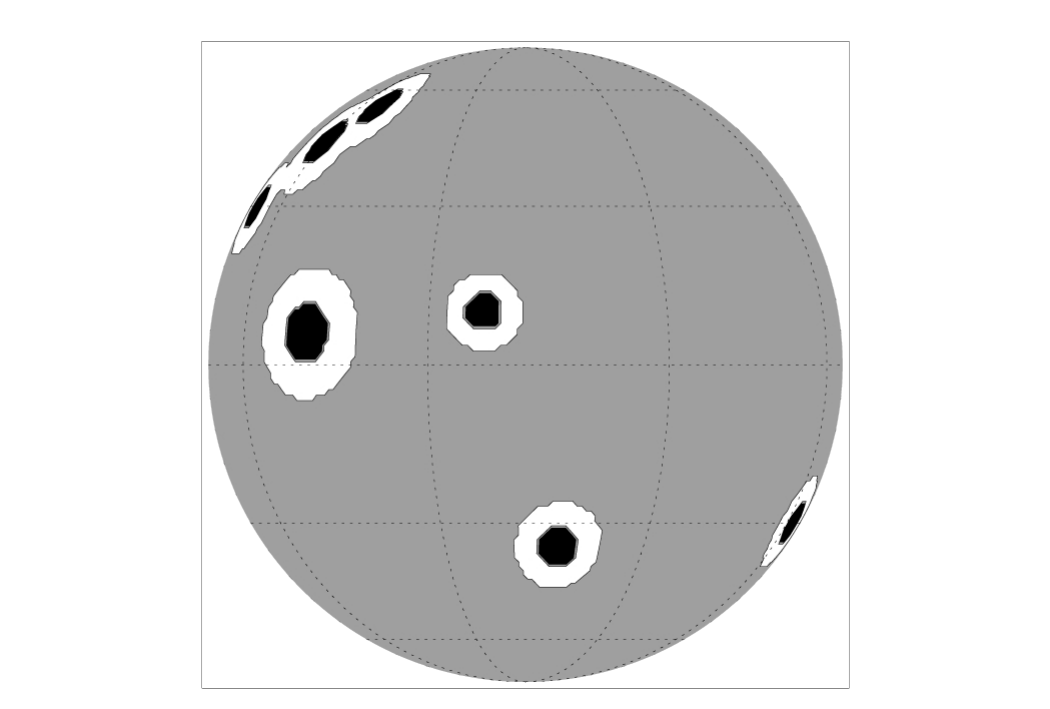}
\end{center}
\caption{Projected map for an arbitrary distribution of active regions with $Q=3.0$. 
The quiet photosphere is represented in grey, and each active region is 
modelled as a cold spot (black) and a hotter surrounding area associated
with faculae (white).
}
\label{fig_map}
\end{figure}

Although the geometric description of the spots is free (only defined by the
properties of the underlying stellar grid), tests have been carried out by
modelling each active region $n$ as a circular spot of radius $R_{\rm Sn}$
surrounded by a corona associated to a facular region, having an external
radius $R_{\rm Fn}=\sqrt{Q+1} \cdot R_{\rm Sn}$ (see Fig.~\ref{fig_map}), where
$Q$ is the ratio between the area of the surrounding faculae and the area of
the spot, which is assumed to be the same for all the active regions on the
star. Our model does not distinguish between spot umbra and penumbra, but
adopts a mean spot contrast at an intermediate temperature, similar to the
prescription used by previous approaches \citep{Chapman87,Unruh99}. The model
of circular spots surrounded by a corona of faculae has been shown to
successfully reproduce spot maps for the Sun \citep{Lanza03} and for several
low mass stars with high precision light curves
\citep{Lanza09b,Lanza10,SilvaValio10}. Solar observations have shown the presence of systematic
changes in the contrast and size of faculae \citep{Chapman87}, with the value
$Q$ decreasing for larger active regions that are more frequent around the
maximum of activity \citep{Chapman11}. \cite{Foukal98} shows that the
relationship between the area of the faculae and the area of the spots is not
linear, and can vary from $Q\sim2$ to $Q\sim10$ depending on the size and
lifetime of the active regions. In our case, we choose values from $Q\sim3$ to
$Q\sim8$, representing medium- and small-sized spots for the simulations of
Sun-like star data. Several studies on the modelling of high precision light
curves conclude that lower values of $Q$ or even $Q=0$ (i.e., no faculae)
provide the best description of the data for K and early M dwarfs
\citep{Gondoin08,Lanza10,Lanza11b}. In our work we investigate and discuss
the appropriate presence of faculae to be included in the simulations of each
particular case.

The geometric distribution of the active regions is not constrained in the
\texttt{StarSim} tool, so this can be either user defined or generated randomly. In
the second case, the longitudes of the active regions are equally distributed
over the stellar surface, whereas the latitudes are distributed in two belts
defined by Gaussian functions centered at $\pm\bar{\theta}_{\rm belt}$. 
This configuration can be used in order to simulate butterfly diagram evolution
patterns, which are  well known for the Sun \citep{Jiang11,Veccio12} and
several active dwarfs \citep{Katsova03,Berdyugina07,Livshits03}, together with
other long term variations \citep{Pulkkinen99}. Latitudes of spots have also
been measured for several stars from observations of spot crossing events
\citep{Sanchis11a}.


Active region evolution can also be modelled with \texttt{StarSim}. This is done by considering a linear increase of the spot 
sizes with time, followed by a time interval of constant size and a final 
decay. In the case of selecting a random distribution of the parameters of
the active regions, the total typical lifetime of active regions,
$\bar{L}_{\rm n}$, its standard deviation, $\sigma (L_{\rm n})$, and the
growing/decaying rate, $a_{\rm rd}$, are also specified when configuring
the simulation.
Spots are known to live for weeks on main sequence 
stars, and months in the case of locked binary systems 
\citep{Hussain02,Strassmeier94a,Strassmeier94b}. However, polar 
starspots have been observed to last for years \citep{Olah91}. 
Measurements of the decay rate of sunspots have shown that it follows a 
parabolic area decay law \citep{Petrovay97,Petrovay99} although previous 
studies assumed both linear and non-linear laws 
\citep{MartinezPillet93}. Regarding other stars, Doppler imaging 
observations of isolated starspots over time are not yet available in 
high resolution. In the case of our simulations, a parabolic law is 
considered, both for the decay and for the emergence of the active 
regions, and the evolution rate is set to the solar values except for 
stars where the surface map has been modelled for a long enough time 
span to estimate spots lifetimes and observe variations in size.

Note that, in the current implementation, our simulator is not intended to model each
active region as a group of small spots, but to consider it as a single circular spot.
We assume that this approach, by considering the appropriate size and evolution
of the active regions, generates mostly the same effects on the simulated 
photometric and spectroscopic data as using a more complicated 
group-pattern configuration. In our simulations, we adjust the number 
and size of active regions to be in agreement with the results 
from the literature that assume a two-temperature structure 
\citep{Henry95,Rodono00,Padmakar99}. \cite{Solanki04} discuss the 
distribution of spot sizes for some Sun-like stars observed with the 
Doppler imaging technique. Although it is suggested that sizes could 
follow a lognormal distribution as on the Sun, we do not adopt such 
distribution because the smaller spots would not cause a significant 
effect on our simulations compared to large spots, especially for the 
most active stars where large spots dominate
\citep{Solanki04,Jackson13,Strassmeier09}.

Spot temperatures have only been determined in a few cases, and the 
inhomogeneity of the results due to the diversity of techniques make 
this still an unconstrained parameter. Most of the current measurements 
made in main sequence stars come from Doppler imaging 
\citep{Marsden05,Strassmeier98} or multi-band light curve modelling 
\citep{Petrov94,ONeal04}. The first ones are limited to young fast-rotating active stars that can present a different kind of activity pattern \citep{Mullan01}, and it is not clear that these results can be extrapolated to the active regions in Sun-like stars. \cite{Berdyugina05} showed that, on average, 
the spot temperature contrast $\Delta T_{\rm spot}$ has a dependence on 
the spectral type, being larger for hotter stars. However, there seems 
to be no correlation between spot temperatures and sizes 
\citep{Bouvier89,Strassmeier92} or activity cycles 
\citep{Stix02,Albregtsen81,Penn07} in the case of the Sun. In our simulations 
we assume the same $\Delta T_{\rm spot}$ for all the active regions, which is estimated from a low order polynomial fit to the data in \cite{Berdyugina05},

\begin{equation}
\Delta T_{\rm spot}=7.9\cdot10^{-5} T_{\rm eff}-0.1056 T_{\rm eff}-153.6.
\label{dtcorr}
\end{equation}

This is consistent with the temperature measurements of the darkest regions of the spot in the case of the Sun \citep{Eker03}. When
possible, we use more accurate determinations of the whole spot region properties 
based on their spectral signature (see Sect.~\ref{specsigna}; also
\citealt{Pont08,Pont13,Sing11}).

As explained in Sect.~\ref{phoenix}, Phoenix models are used to 
reproduce the spectral signature of faculae in our approach. Faculae are 
known to be mainly a magnetic phenomenon affecting the intensity of the 
spectral lines, best reproduced by NLTE simulations \citep{Carlsson92}. 
The lack of high resolution models for faculae in the spectral range of 
our interest prevents us from including them into high resolution spectra 
simulations, restricting them to the simulations of broad band 
photometric variations (see Sect.~\ref{sim_photo}). A positive temperature 
contrast between the faculae and the photosphere, $\Delta T_{\rm 
fac}\simeq 30-50$~K is assumed, as reported by \cite{Badalyan73} from 
observations of the CO line in faculae and by \cite{Livshits68} from a 
more theoretical point of view. \cite{Solanki93} provides an 
extensive discussion and references on the weakening of lines in faculae 
and the needed $\Delta T_{\rm fac}$ to best reproduce the continuum 
brightness.

\begin{table}
\caption{Input parameters}
\label{tab:simparams}
\begin{tabular}{ll}
\hline\noalign{\smallskip}
  Parameter & Typical values \\
\noalign{\smallskip}\hline\noalign{\smallskip}
\hline
  Initial time of simulation (days) & $-$ \\
  Final time of simulation (days)  & $-$ \\
  Data time cadence (minutes) & $-$ \\
  Working mode & Phot. / Spec. \\
  Spectral range (nm) &  500 $-$ 50000 \\
  Photometric filter & $-$ \\
  Grid resolution ($^{\circ}$) & $1^{\circ}\times1^{\circ}$ \\
  RV window of the & \\
  \hspace{2mm}CCF, $\Delta v_{\rm CCF}$ (km s$^{-1}$) & 20 $-$ 40  \\
  \hline
  Add photon noise & y/n \\
  Stellar apparent magnitude $K$ & $-$ \\
  Telescope area (m$^{2}$) & $-$ \\
  Instrument efficiency & 0 $-$ 1 \\
  \hline
  $T_{\rm eff}$ (K) & 2600 $-$ 12000 \\
  $\log g$ & +3.5 $-$ +5.0 \\
  $\rm[Fe/H]$ & $-$4.0 $-$ +1.0 \\
  $\rm[\alpha/Fe]$ & 0.2 $-$ +1.2 \\
  Spots temperature contrast, & \\
   $\Delta T_{\rm spot}$ (K) & 200 $-$ 1500\\
  Faculae temperature contrast, & \\
   $\Delta T_{\rm fac}$ (K) & 30 $-$ 50 \\
  Facula-to-spot area ratio, $Q$ & 0.0 $-$ 10.0 \\
  Stellar rotation period, $P_{0}$ (days) & $-$ \\
  Differential rotation, $k_{\rm rot}$ & $-$ \\
  Inclination angle, $i_{*}$ ($^{\circ}$) & 0 $-$ 90$^{\circ}$ \\
  \hline
  Minimum number of spots, $Nr_{min}$ & $-$ \\
  Maximum number of spots $Nr_{max}$ & $-$ \\
  Mean area of spots $\bar{A}_{\rm Sn}$ & $-$ \\
  Standard deviation of the area of & \\
   spots $\sigma(A_{\rm Sn})$ & $-$ \\
  Mean lifetime of active regions, & \\
   $\bar{L}_{\rm n}$ (days) & $10^{1} $-$ 10^{3}$ \\
  Standard deviation of lifetime of & \\
   active regions $\sigma (L_{\rm n})$ (days) & $-$ \\
  Growing/decaying rate $a_{\rm rd}$ ($^{\circ}$/day) & $-$ \\
  Mean latitude of active regions & \\
   $\pm\bar{\theta}_{\rm belt}$ ($^{\circ}$) & 0 $-$ 90$^{\circ}$ \\
  Standard deviation of latitude of & \\
   active regions $\sigma(\theta_{\rm belt})$ ($^{\circ}$) & 0 $-$ 90$^{\circ}$ \\
  \hline
  Planet radius, $R_{\rm planet}/R_{*}$ & $-$ \\
  Mid transit time, $T_{0}$ (days) & $-$ \\
  Orbital Period, $P_{\rm planet}$ (days) & $-$ \\
  Impact parameter, $b$ & 0 $-$ 1 \\
  Spin-orbit angle, $\gamma$ ($^{\circ}$) & 0 $-$ 90$^{\circ}$ \\
\noalign{\smallskip}\hline
\end{tabular}
\end{table}

A transiting planet can also be introduced in the simulations to investigate
activity effects on photometric and radial 
velocity observations during transits. The planet is modelled as a 
circular black disk and is described in the same way as a spot with zero 
flux, but assuming a circular projection on the stellar disk. No planetary
atmosphere is considered in the current implementation and so the radius 
of the planet has no wavelength dependence. The photometric signal for the 
primary transit is generated from the planet size, the ephemeris and the 
orbit orientation (inclination and spin-orbit angles), which are 
specified as input parameters. The orbital semi-major axis is computed 
from the orbital period, $P_{\rm planet}$, and the stellar mass, 
$M_{*}$, using Kepler's Third Law by assuming $M_{*} \gg m_{\rm 
planet}$. Eccentricity is considered to be zero for all cases in the 
current model.

Together with the physical parameters of the star-planet system, the 
spectral range for the output data, the timespan of the simulations and the
cadence of the data series can be selected. The program generates the 
resulting time series spectra and also creates a light curve by 
multiplying each generated spectrum by a filter passband specified from a
database. If preferred, a filter with a rectangular transfer function 
can be defined and used to study the photometric signal at any 
desired wavelength range. Finally, the integration time, the telescope
collecting area, the efficiency of the instrument and the target star
magnitude are needed to apply photon noise statistics to the resulting fluxes.

\subsubsection{Simulating photometric time series}
\label{sim_photo}

To build the spectrum of the spotted stellar surface, three initial 
synthetic spectra generated from models with different temperatures are 
computed for the quiet photosphere, spots and faculae. The program 
reads the physical parameters for the three photospheric features of the 
modelled star and interpolates in $T_{\rm eff}$ within the corresponding 
model grids to generate the three synthetic spectra: 
$f_{p}(\lambda),f_{s}(\lambda)$ and $f_{f}(\lambda)$. The rest of the 
stellar parameters (i.e. $\log g$, metallicity, etc.) are not 
significantly variable over the stellar surface, and specifically not for 
active regions. Therefore, their precision is not critical for the 
purpose of our simulations, and the nearest grid values are 
considered instead of interpolations.

Kurucz (ATLAS9) spectra are computed for the specified parameters
reproducing the photosphere and the spots. These models provide information on
the intensity profile of the star, which can be used in order to compute the
limb darkening factors, $I(\lambda,\mu)/I(\lambda,0)$.  With our approach, such a
 factor is computed separately for every surface element and
wavelength. This is done by interpolating within the grid of intensities
provided by the Kurucz models. Then, the spectrum of the undarkened surface
element ($f_{p}(\lambda)$ for the quiet photosphere or $f_{s}(\lambda)$ for the
spots) is multiplied by the corresponding limb darkening factor,
$I_{p}(\lambda,\mu)/I_{p}(\lambda,0)$ or $I_{s}(\lambda,\mu)/I_{s}(\lambda,0)$
(see Eqs.~\ref{eqflim}, \ref{eqdeficit}, \ref{eqtrans}). In the case of
faculae, a different approach is considered, as it is known that these areas
are brightened near the limb \citep{Frazier78,Berger07}. A limb brightening law
is considered in this case, as presented in Eq.~\ref{lbright}.

The code produces time series data considering the 
initial and final times and the cadence of the time series specified in 
the input parameters. For each time step, the evolutionary stage 
(current size) and the projected position of each active region are 
computed. This last parameter is obtained from the specified rotation period, 
$P_{0}$, and the differential rotation of the star, which is modelled as 
described by \cite{Beck00}:

\begin{equation}
\Omega(\theta)=\Omega_{0}+k_{\rm rot}(\Omega_{\odot \rm A} \cdot \cos^{2}\theta+\Omega_{\odot \rm B} \cdot \cos^{4}\theta),
\label{eqprot}
\end{equation}

\noindent where $\Omega_{0}$ is the equatorial rate, $\Omega_{\odot \rm 
A}=1.698^{\circ}$day$^{-1}$ and $\Omega_{\odot \rm B}=2.346^{\circ}$day$^{-1}$ are 
the coefficients that describe the differential rotation for the Sun 
\citep{Snodgrass90}, $k_{\rm rot}$ is a factor that sets the 
differential rotation rate for the star and is specified in the input 
parameters (the value for the Sun would be $k_{\rm rot}$=1), and 
$\theta$ is the colatitude.

In a first step, the spectrum for the quiet photosphere is obtained 
from the contribution of all the surface elements, taking into account 
the geometry of the element, its projection towards the observer, the 
corresponding limb darkening profile and the RV shift:

\begin{eqnarray}
f_{im}(\lambda) &=& \sum\limits_{k} f_{p}(\lambda_{k},\mu_{k},a_{k})=\nonumber\\
&=& \sum\limits_{k} f_{p}(\lambda_{k}) \cdot \frac{I_{p}(\lambda_{k},\mu_{k})}{I_{p}(\lambda_{k},0)}\cdot a_{k} \cdot \mu_{k} \cdot \omega_{k},
\label{eqflim} 
\end{eqnarray}

\noindent where $k$ is the surface element and $\mu_{k}$ is the cosine 
of its projection angle given by

\begin{equation}
\mu_{k}=\sin i_{*} \sin \theta_{k} \cos \phi_{k} + \cos i_{*} \cos \theta_{k},
\label{musimp}
\end{equation}

\noindent where $\theta_{k}$ and $\phi_{k}$ are the colatitude and 
longitude coordinates, respectively, and $i_{*}$ is the stellar axis 
inclination.

The factor $\omega_{k}$ accounts for the visibility of the surface 
element and is given by

\begin{equation}
\omega_{k}=\left\{\begin{array}{rl}
1 &\mbox{ if $\mu_{k} \ge 0$} \\
0 &\mbox{ if $\mu_{k} < 0$,}
\end{array} \right.
\label{eqome} 
\end{equation}

\noindent and $a_{k}$ is the area of the surface element, which can be 
computed from the spatial resolution of the grid ($\Delta\alpha$) by

\begin{equation}
a_{k}=2 \cdot \Delta\alpha \cdot \sin (\frac{\Delta\alpha}{2})\sin\theta_{k}.
\label{are}
\end{equation}

Finally, $\lambda_{k}$ are the wavelengths including the Doppler shift 
for the corresponding surface element,

\begin{equation}
\lambda_{k}=\lambda+\Delta\lambda_{k},
\label{rvshifted}
\end{equation}

\noindent with

\begin{equation}
\Delta\lambda_{k}\simeq-8.05\cdot\lambda\cdot\frac{1}{c}\cdot R_{\rm star} \cdot \frac{2\pi}{P_{0}}\sin i_{*} \sin\theta_{k} \sin\phi_{k}.
\label{rvshift}
\end{equation}

These are computed from the equatorial rotation period given in the 
input parameters and the stellar radius calculated using its relation 
with $\log g$ and $T_{\rm eff}$ for main sequence stars.

The flux variations produced by the visible active regions are added to 
the contribution of the immaculate photosphere when computing the 
spectrum at each time step $j$. These are given by

\begin{eqnarray}
&& \Delta f^{ar}_{j}(\lambda) = \nonumber\\
&& = \sum\limits_{k} [(f_{s}(\lambda_{k}) \cdot J^{kj}_{s}(\lambda_{k})-f_{p}(\lambda_{k}) \cdot J^{kj}_{p}(\lambda_{k})) \cdot p^{kj}_{s} + \nonumber \\
&&+ (f_{f}(\lambda_{k}) \cdot J^{kj}_{f}-f_{p}(\lambda_{k}) \cdot J^{kj}_{p}(\lambda_{k}))  \cdot p^{kj}_{f}],
\label{eqdeficit} 
\end{eqnarray}

\noindent where the first term accounts for the flux deficit produced by 
the spots and the second is the overflux produced by faculae. The 
quantities $J^{kj}_{p}(\lambda_{k})$, $J^{kj}_{s}(\lambda_{k})$ and 
$J^{kj}_{f}$ account for the geometric and the limb 
darkening/brightening factors for the surface element $k$ at the time 
step $j$ of the simulation, and are given by:

\begin{eqnarray}
&& J^{kj}_{p}(\lambda_{k})= \frac{I_{p}(\lambda_{k},\mu_{kj})}{I_{p}(\lambda_{k},0)}\cdot a_{k} \cdot \mu_{kj} \cdot \omega_{kj}\nonumber\\
&& J^{kj}_{s}(\lambda_{k})= \frac{I_{s}(\lambda_{k},\mu_{kj})}{I_{s}(\lambda_{k},0)}\cdot a_{k} \cdot \mu_{kj} \cdot \omega_{kj} \nonumber\\
&& J^{kj}_{f}= c_{\rm f}(\mu_{kj}) \cdot a_{k} \cdot \mu_{kj} \cdot \omega_{kj}.
\label{eqdefintens} 
\end{eqnarray}

The factors $p^{kj}_{s}$ and $p^{kj}_{f}$ are the fractions of the 
surface element $k$ covered by spot and faculae, respectively. These 
amounts are computed for every surface element considering their 
distance to the center of all the neighbour active regions at each 
observation $j$. There is also a time dependence of the projection of 
the surface elements, given by

\begin{equation}
\mu_{kj}=\sin i_{*} \sin \theta_{k} \cos [\phi_{k} + \Omega(\theta) \cdot (t_{j}-t_{0})] + \cos i_{*} \cos \theta_{k}.
\label{mu}
\end{equation}

Finally, $c_{f}(\mu_{kj})$ accounts for the limb brightening of faculae, 
which is assumed to follow the law used by \cite{Meunier10}:

\begin{equation}
c_{\rm f}(\mu_{kj})=\left(\frac{T_{\rm eff}+\Delta T_{\mu}(\mu_{k})}{T_{\rm eff}+\Delta T_{\rm fac}}\right)^{4},
\label{lbright}
\end{equation}

\noindent where $\Delta T_{\mu}(\mu)=a_{\mu}+b_{\mu} \cdot \mu_{k} + c_{\mu}\cdot \mu_{k}^{2}$.


The coefficients are $a_{\mu}=250.9$, $b_{\mu}=-407.4$ and 
$c_{\mu}=190.9$, so that $c_{\rm 
f}(\mu_{kj})\sim1$ at the center of the disk and $\sim1.16$ near the 
limb for a Sun-like star with $\Delta T_{\rm fac}=30$~K. This is in 
agreement with the parametrizations presented by \cite{Unruh99}. The 
observations show that the contribution of faculae dominate the 
irradiance in the case of the Sun, where active regions are $\sim$1.2 
times brighter at the limb than at the center of the stellar disk 
\citep{Ortiz02,Ball11,Steinegger96}.

As explained in Sect.~\ref{params}, a transiting planet can be included 
 as a dark circular spot with a constant radius over 
wavelength (no atmosphere) crossing the stellar disk. For each time step 
$j$, the planet position is obtained from the given ephemeris and the 
flux deficit is computed from the eclipsed surface elements of the star:

\begin{equation}
\Delta f^{tr}_{j}(\lambda) = -\sum\limits_{k} f_{p}(\lambda_{k}) \cdot J^{kj}_{p}(\lambda_{k}) \cdot p^{kj}_{tr},
\label{eqtrans} 
\end{equation}

\noindent where $p^{kj}_{tr}$ is the fraction of the surface element $k$ 
covered by the planet at time step $j$. In case an active 
region is partially occulted by the planet, the corresponding spot 
($p^{kj}_{s}$) and facula ($p^{kj}_{f}$) fractions (see 
Eq.~\ref{eqdeficit}) in the occulted surface elements are computed 
within the planet covered fraction ($p^{kj}_{tr}$) instead.

Finally, the spectrum for the observation $j$ is obtained by adding the 
contribution of the immaculate photosphere, the active regions and the 
transiting planet:

\begin{equation}
f_{j}(\lambda)=f^{im}(\lambda) + \Delta f^{ar}_{j}(\lambda) + \Delta f^{tr}_{j}(\lambda).
\label{spectot} 
\end{equation}

When working with the low resolution Phoenix spectra to produce a
photometric time series, the resulting spectra are multiplied by the 
specified filter passband (see Fig.\ref{fig_sun}) to obtain the total 
flux and to compute the jitter produced by activity at the desired spectral 
range. The photometric jitter is evaluated by computing the rms of the 
flux over time for the whole photometric data series.

\begin{figure}[]
\begin{center}
\includegraphics[width=\columnwidth,angle=0]{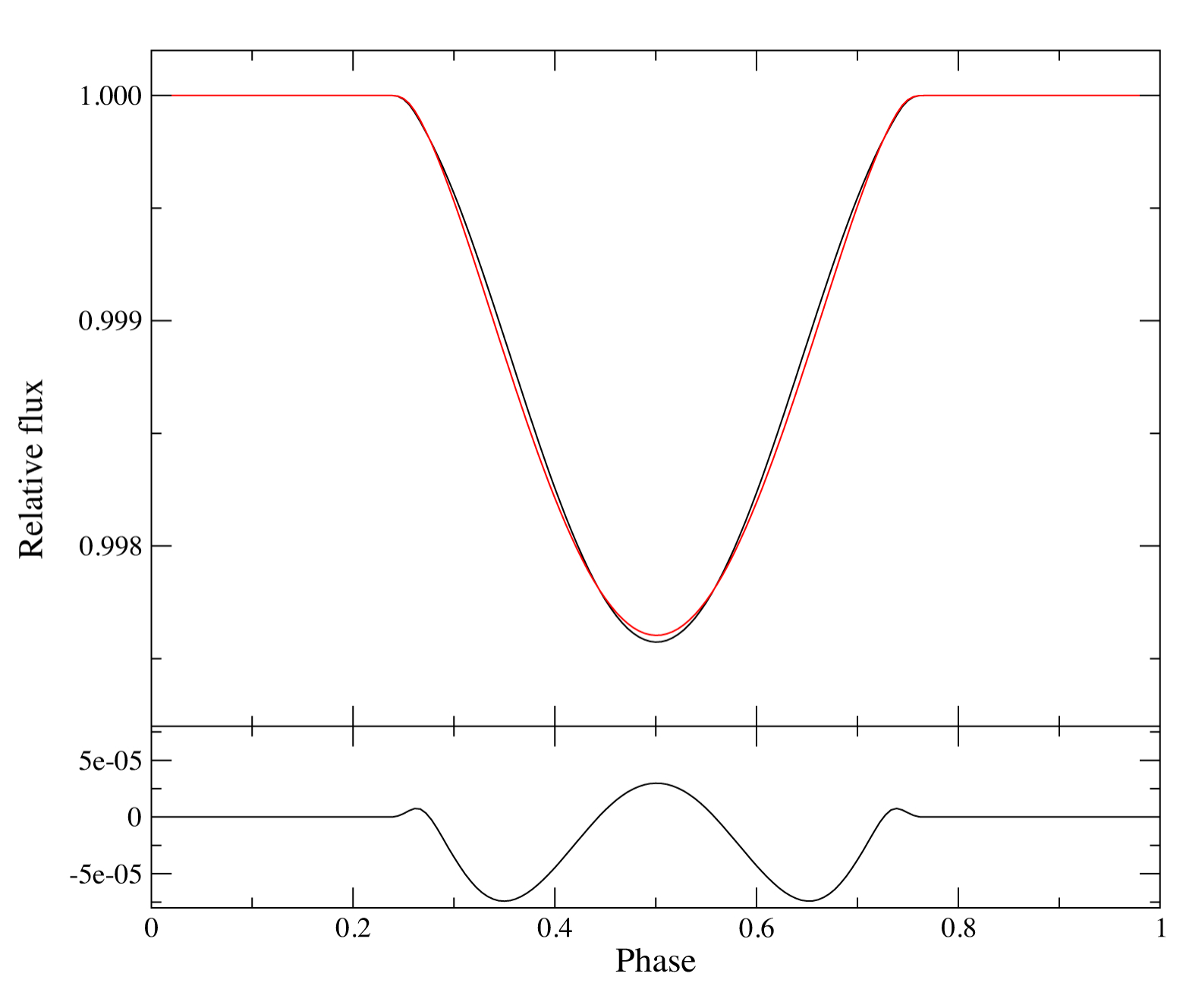}
\end{center}
\caption{Top panel: A comparison between two light curves obtained for a
star with $T_{\rm eff}=5777$~K, $\log g=4.5$ and  $\Delta T_{\rm spot}=320$~K
from Phoenix BT-Settl models (red) and solar observed spectra (black), by
introducing an equatorial spot covering $\sim0.3\%$ of the visible stellar
surface.
Bottom panel: The difference between both light curves.  }
\label{suncomp}
\end{figure}

A comparison between two light curves obtained from Phoenix models and 
solar observed spectra for the Johnson-V band 
is presented in Fig.~\ref{suncomp}. The simulation covers one rotation
period of a Sun-like star with a single circular spot and no faculae. The spot
is located on the equator and covers $\sim0.3\%$ of the visible stellar
surface. The temperature contrast of the spot, $\Delta T_{\rm spot}$, was
fitted considering 10~K steps in the Phoenix spectrum of the spot and minimizing the rms of the residuals between
both light curves. The best solution was found for $\Delta T_{\rm spot}=320$~K,
presenting differences at the $10^{-5}$ level, which is in close agreement with
the contrast found by direct determination from the spectra (see
Sect.~\ref{mod_sol}). We conclude that the small deviations shown by the
models for specific spectral lines do not have a significant impact at the
required photometric level when considering broad spectral bands.


The position of the photocenter on the stellar disk is also computed for 
each time step $j$ of the simulations. This is implemented for a further 
investigation of the astrometric jitter caused by the presence of active 
regions or transiting planets. The astrometric shifts depend on wavelength and are computed 
separately for the equatorial axis ($X$) and the spin projected axis of 
the star ($Y$) by:

\begin{eqnarray}
&& \Delta X^{j}(\lambda) = \sum\limits_{k} \{f_{p}(\lambda_{k}) \cdot J^{kj}_{p}(\lambda_{k})\cdot x_{kj} +\nonumber\\
&& + [f_{s}(\lambda_{k}) \cdot J^{kj}_{s}(\lambda_{k})-f_{p}(\lambda_{k})\cdot J^{kj}_{p}(\lambda_{k})] \cdot p^{kj}_{s} \cdot x_{kj} + \nonumber\\
&& + [f_{f}(\lambda_{k})\cdot J^{kj}_{f}-f_{p}(\lambda_{k})\cdot J^{kj}_{p}(\lambda_{k})]\cdot p^{kj}_{f}\cdot x_{kj} + \nonumber \\
&& + f_{p}(\lambda_{k}) \cdot J^{kj}_{p}(\lambda_{k}) \cdot p^{kj}_{tr} \cdot x_{kj}\}
\label{astroshX} 
\end{eqnarray}

\noindent and

\begin{eqnarray}
&& \Delta Y^{j}(\lambda) = \sum\limits_{k} \{f_{p}(\lambda_{k}) \cdot J^{kj}_{p}(\lambda_{k})\cdot y_{kj} +\nonumber\\
&& + [f_{s}(\lambda_{k}) \cdot J^{kj}_{s}(\lambda_{k})-f_{p}(\lambda_{k})\cdot J^{kj}_{p}(\lambda_{k})] \cdot p^{kj}_{s} \cdot y_{kj} + \nonumber\\
&& + [f_{f}(\lambda_{k})\cdot J^{kj}_{f}-f_{p}(\lambda_{k})\cdot J^{kj}_{p}(\lambda_{k})]\cdot p^{kj}_{f}\cdot y_{kj} + \nonumber \\
&& + f_{p}(\lambda_{k}) \cdot J^{kj}_{p}(\lambda_{k}) \cdot p^{kj}_{tr} \cdot y_{kj}\},
\label{astroshY} 
\end{eqnarray}

\noindent where the quantities $J^{kj}_{p}(\lambda)$, 
$J^{kj}_{s}(\lambda)$ and $J^{kj}_{f}(\lambda)$ are given by 
Eq.~\ref{eqdefintens} and $x_{kj}$ and $y_{kj}$ denote the position of 
the surface element $k$ projected on the $X$ and $Y$ axes, respectively, 
at the time step $j$ of the simulation. The rest of the variables are 
defined in Eqs.~\ref{eqflim} to \ref{eqtrans}.

The first term in Eqs.~\ref{astroshX} and \ref{astroshY} provides the 
position of the photocenter for an immaculate photosphere, whereas the 
second, third and fourth add the contribution of spots, faculae and 
planet to the astrometric shifts. As in Eq.~\ref{eqtrans}, the 
corresponding flux contribution is computed within $p^{kj}_{tr}$ and 
subtracted from $p^{kj}_{s}$ or $p^{kj}_{f}$, when the planet is 
occulting a surface element $k$ containing a spot or a facula, 
respectively.

\begin{figure*}
\centering
    \resizebox{0.95\hsize}{!}{\includegraphics{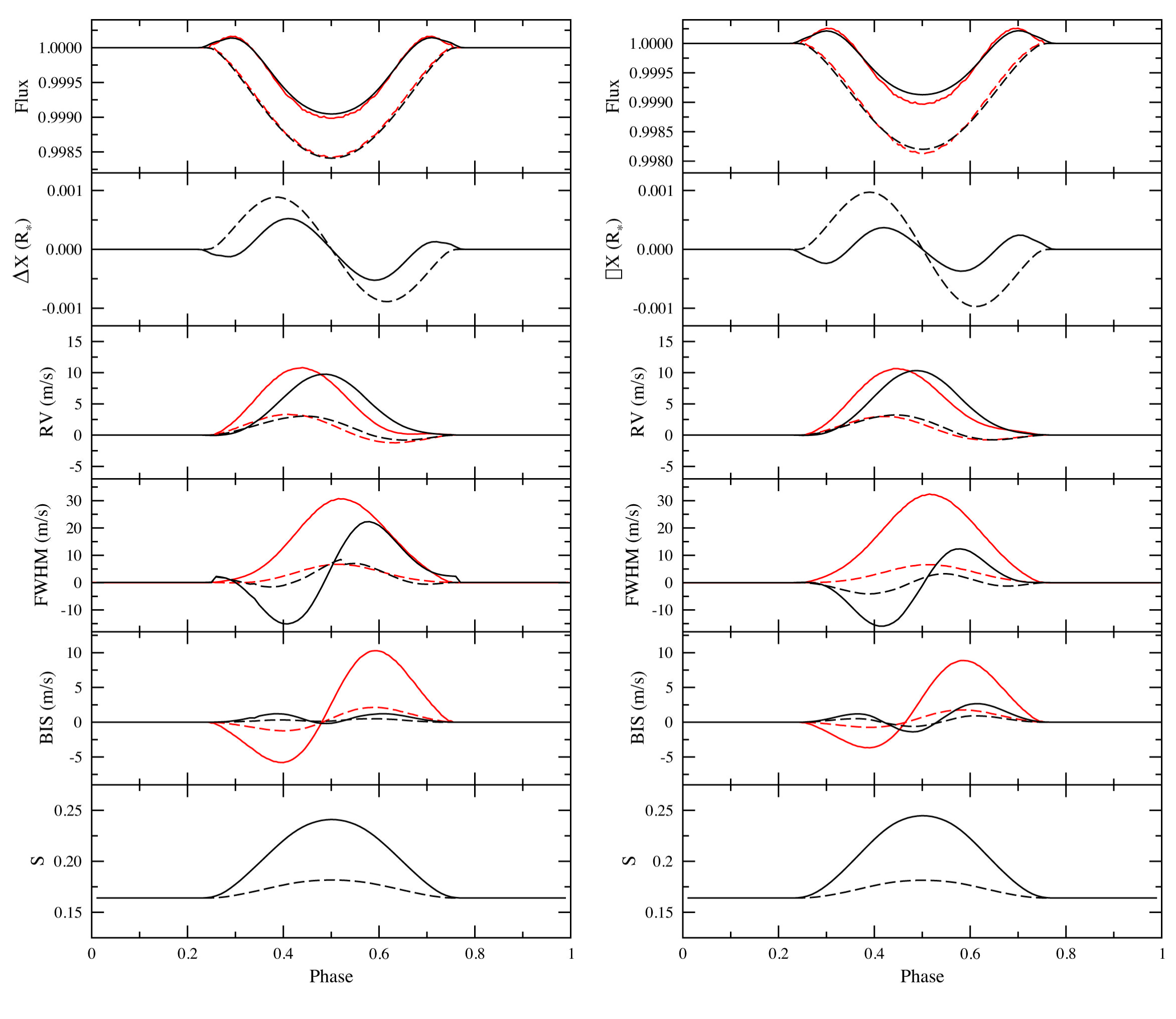}}
    \caption{
Simulations of an active region located on the equator $\theta=90^{\circ}$
with a size of ${A}_{\rm Sn}=1.6 \cdot 10^{-3}$ relative to the stellar
surface.
The rotation period is 25 days and the inclination is $90^{\circ}$, yielding
$v\sin i = 2$ km s$^{-1}$. Left panels: simulated data for a Sun-like star
($T_{\rm eff}=5770$~K) and $\Delta T_{\rm spot}=310$~K. Right panels: the same
for a K0 star ($T_{\rm eff}=5000$~K) and $\Delta T_{\rm spot}=250$~K. From top
to bottom: photometric (in Johnson V band), astrometric, RV, FWHM, BIS and S
index variations induced by the active region. The results with surrounding
faculae ($Q=3$) are shown with black solid lines and the ones with no faculae
($Q=0$) are plotted with black dashed lines. The same simulations made with
SOAP 2.0 are plotted with red lines.
}
\label{simjitters}
\end{figure*}

$\Delta X^{j}(\lambda)$ and $\Delta Y^{j}(\lambda)$ vectors are finally 
multiplied by a given instrumental response function or a filter 
passband to produce astrometry displacement curves. Two simple cases are
presented in Fig.~\ref{simjitters} for a Sun-like and a K0-type star observed
at $i=90^{\circ}$. A single circular active region was considered, consisting
of a spot of size ${A}_{\rm Sn}=1.6 \cdot 10^{-3}$ relative to the stellar
surface and a surrounding facular area with $Q=3.0$, located on the equator of
the star ($\theta=90^{\circ}$). The temperature contrast of the spot was set to
$\Delta T_{\rm spot}=310$~K for the Sun-like star and  $\Delta T_{\rm
spot}=250$~K for the K0 star (see Sect.~\ref{params} and
\citealt{Berdyugina05}). The photometric signature can be seen in the top
panel, which is dominated by faculae when the active region appears at
the edges of the stellar disk (phases $\sim0.3$ and $\sim0.7$) as an effect of
limb brightening, while the dark spot produces a decrease of $\sim0.001$ in
relative flux units at the center of the stellar disk (phase$\sim$0.5). The astrometric
signature (middle panel) peaks at $\pm0.001 R_{*}$. This jitter would 
only affect astrometric measurements at the expected precision of the 
Gaia space mission \citep{Lindegren08,Eyer13} for main sequence stars located 
at 20~pc or closer. Further analysis and discussion on astrometric 
jitter simulations will be presented in a future paper.

\subsubsection{Simulating spectroscopic data}
\label{sim_rv}

A similar methodology as the one presented in Sect.~\ref{sim_photo} is 
implemented for the simulation of time series data to obtain 
activity-induced RV curves. A drawback of the methodology 
used is the large volume of data produced when working with high-resolution
spectra, which is solved by working in the velocity space, as explained below. 

Three synthetic spectra ($f_{p}(\lambda),f_{s}(\lambda)$ and 
$f_{f}(\lambda)$) for the given wavelength range are computed for 
different temperatures corresponding to the quiet photosphere, spots and 
faculae, by interpolating from \cite{Husser13} Phoenix library (see 
Sect.~\ref{phoenix}). The nearest values in the grid to the specified 
ones are considered for the rest of the parameters, as done in 
Sect.~\ref{sim_photo}. Then, individual CCFs 
($C_{p}(v),C_{s}(v)$ and $C_{f}(v)$) are computed, respectively, from 
the three spectra and a template mask consisting of a synthetic spectrum 
of a slowly rotating inactive star of the same parameters. In the current version of \texttt{StarSim},
the mask templates from the HARPS instrument pipeline (DRS,\citealt{Cosentino12}) are used in order to compute the CCFs contribution of the different elements \citep{Pepe02,Pepe04}. These are optimized for G2, K5 and M2 type main sequence stars and consist on a selection of delta-peak lines, whose height corresponds to the line equivalent width.

The global CCF of the projected stellar surface is computed from the 
contribution of the individual surface elements. First, the signature of 
the immaculate photosphere is obtained from

\begin{equation}
C^{im}(v) = \sum\limits_{k} C_{p}(v_{k},\mu_{k},a_{k})= \sum\limits_{k} C_{p}(v_{k}) \cdot H^{k}_{p},
\label{eqCim} 
\end{equation}

\noindent where the quantity $H^{k}_{p}$ accounts for the intensity and 
geometric factors of the surface element $k$,

\begin{equation}
H^{k}_{p}=\sum\limits_{\lambda}f_{p}(\lambda) \cdot \frac{I_{p}(\lambda,\mu_{k})}{I_{p}(\lambda,0)}\cdot a_{k} \cdot \mu_{k} \cdot \omega_{k}.
\label{eqCimins} 
\end{equation}

\noindent The summation covers the whole wavelength range specified for 
$f_{p}(\lambda)$.

The shifted velocities, $v_{k}$, are computed from the contribution of 
the Doppler shift and the effects of convection,

\begin{equation}
v_{k}=v+\Delta v^{\rm DS}_{k} + \Delta v^{\rm C}_{k}(C_{l}),
\label{eqvshift} 
\end{equation}

\noindent for $l=\{p,s,f\}$, where $\Delta v^{\rm DS}_{k}$ is the 
contribution from the Doppler shift for the surface element $k$ (see 
Eq.~\ref{rvshift}) and $\Delta v^{\rm C}_{k}(C_{i}(v))$ is the effect of 
convective blueshift, which depends on the line depth and can be 
characterized by studying the line bisectors, which will typically show 
a distinctive C-shape due to granulation in the photosphere 
\citep{Gray92}.

\begin{figure*}
\centering
    \resizebox{\hsize}{!}{\includegraphics{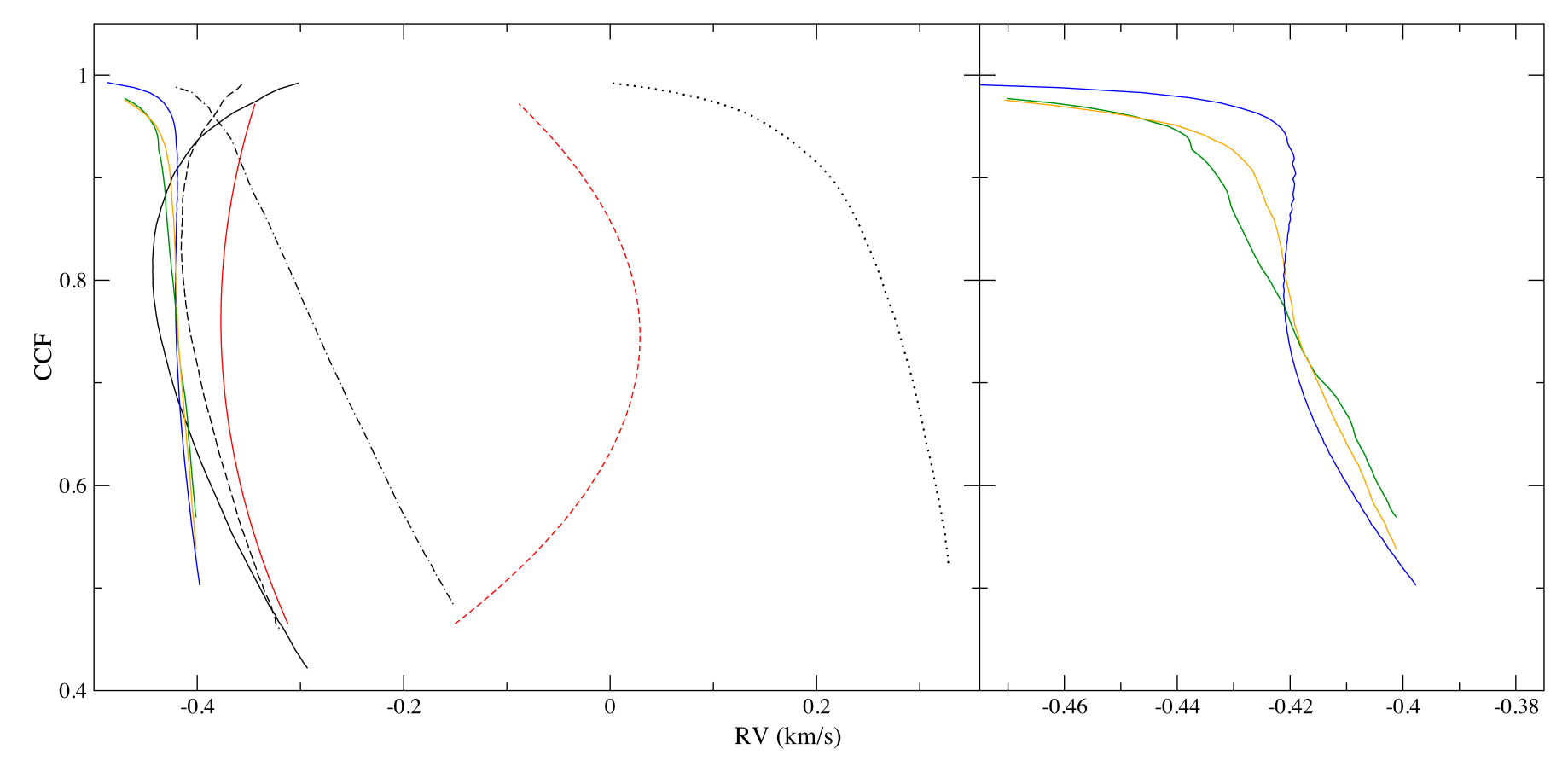}}
    \caption{Left panel: Line bisectors computed using the full HARPS 
instrument range ($\sim$380 - 670 nm) from CIFIST 3D model spectra at 
$\mu=1.0$ (black solid line), $\mu=0.79$ (black dashed line), $\mu=0.41$ 
(black dot-dashed line) and $\mu=0.09$ (black dotted line). Also shown are
line bisectors for the solar observed spectra for a quiet Sun region (red solid
line) and a spot (red dashed line), and the the bisector computed with 
\texttt{StarSim} from the integrated CCF of an unspotted photosphere of a Sun-like star
(blue solid line). Right panel: Zoom of the left part of the left plot showing
a comparison with the bisectors of the Sun-like stars $\alpha$~Cen~A (green solid line) and 18~Sco (orange solid line) observed with the HARPS South instrument (X. Dumusque, private communication).
} 
     \label{fig:bisectors}
\end{figure*}

Note that the effects of convection in the line profile are included in the 
Phoenix spectra from \cite{Husser13} (see Sect.~\ref{starsimmodels}) 
assuming a surface element in LTE observed at $\mu=1$. However, this 
description is too simple and only valid for surface elements located at
the center of the stellar disk. Therefore, we decided to employ a more
sophisticated treatment introducing the convective blueshift from CIFIST 3D models \citep{Ludwig09} including NLTE
effects. The convection effects of the 
Phoenix spectra are fitted with a low order polynomial function 
to the CCF computed with the full spectral range, and then subtracted. 
Next, we add the convective blueshift contribution computed from the 3D models. They are available for a Sun-like star and several projection angles ranging 
from the disk center to the stellar limb (C. Allende Prieto, private communication). We computed the CCFs and then fitted low order polynomial functions to the line bisectors for the different available projection angles. Then, we 
obtained the $\Delta v^{\rm C}_{k}(C_{p})$, $\Delta v^{\rm 
C}_{k}(C_{s})$ and $\Delta v^{\rm C}_{k}(C_{f})$ for any projection 
angle $\mu_{k}$ from a linear interpolation of the available models. In 
the case of the active regions (i.e., both spot and facula zones), 
where convection is known to be blocked by strong magnetic fields (see 
\citealt{Strassmeier09} and references therein), the solar high-resolution 
observed spectra described in Sect.~\ref{mod_sol} are used in 
order to compute the shift in the line bisector with respect to the 
photosphere. This is then scaled with $\mu$ and added to the amounts $\Delta 
v^{\rm C}_{k}(C_{s})$ and $\Delta v^{\rm C}_{k}(C_{f})$ of the surface 
elements where a contribution of spot or faculae is present. In this way, the contribution of active regions is redshifted $\sim0.3$~km~s$^{-1}$ (see Fig.~\ref{fig:bisectors}) due to the inhibition of convection \citep{Dumusque14a}.

 Figure~\ref{fig:bisectors} shows four of the bisectors computed by using 
the full HARPS wavelength range ($\sim$380 - 670 nm) from CIFIST 3D 
model spectra and by using the solar observed spectra for a quiet Sun 
region and a spot. The right panel also 
shows a comparison with the bisectors of 18~Sco and $\alpha$~Cen~A, which have very similar physical properties as the Sun and were observed with
HARPS at a minimum activity level (X. Dumusque, priv. comm.). They are displayed together with the simulation of the bisector of a spotless photosphere of a Sun-like star generated with \texttt{StarSim}. The differences are  below
$\sim$10-20~m~s$^{-1}$ and $<$10~m~s$^{-1}$ for contrasts $<$0.9, so
that the simulated bisectors reproduce the general behavior of the HARPS
observations. The differences at the top of the bisector
could be due to deviations from the real limb darkening profile, i.e. having
more weight at the stellar disk center would increase the effect of C-shaped
lines. Also, more accurate modelling of convection effects on shallow lines
would be needed in order to improve the agreement on this part of the
bisector.

The vertical component of the convective velocity produces a maximum 
shift at the center of the stellar disk, while the effect is barely seen 
near the limb where the projected velocity is zero. On the other hand, 
an apparent redshift of the spectral lines in comparison to the quiet 
photosphere is observed in spotted and facular areas as convective 
motions are blocked \citep{Meunier10b}. Therefore, convection effects 
will introduce a maximum shift in RV measurements when a 
dark spot is crossing the center of the stellar disk.

The variations produced by active regions on the observed CCF at each 
time step $j$ of the simulation are given by

\begin{eqnarray}
&& \Delta C^{ar}_{j}(v) = \nonumber\\
&& = \sum\limits_{k} \{[C_{s}(v_{k}) \cdot H^{kj}_{s}-C_{p}(v_{k}) \cdot H^{kj}_{p}] \cdot p^{kj}_{s} + \nonumber \\
&& + [C_{f}(v_{k}) \cdot H^{kj}_{f}-C_{p}(v_{k}) \cdot H^{kj}_{p}] \cdot p^{kj}_{f}\},
\label{eqCdeficit} 
\end{eqnarray}

\noindent where the quantities $H^{k}_{p}$, $H^{k}_{s}$ and $H^{k}_{f}$ 
are computed by

\begin{eqnarray}
&& H^{kj}_{p}=\sum\limits_{\lambda}f_{p}(\lambda) \cdot \frac{I_{p}(\lambda,\mu_{kj})}{I_{p}(\lambda,0)}\cdot a_{k} \cdot \mu_{kj} \cdot \omega_{kj} \nonumber\\
&& H^{kj}_{s}=\sum\limits_{\lambda}f_{s}(\lambda) \cdot \frac{I_{s}(\lambda,\mu_{kj})}{I_{s}(\lambda,0)}\cdot a_{k} \cdot \mu_{kj} \cdot \omega_{kj} \nonumber\\
&& H^{kj}_{f}=\sum\limits_{\lambda}f_{f}(\lambda)\cdot c_{f}(\mu_{kj}) \cdot a_{k} \cdot \mu_{kj} \cdot \omega_{kj}.
\label{eqCimins2} 
\end{eqnarray}

The first term in Eq.~\ref{eqCdeficit} provides the contribution of 
spots, while the second is the contribution for the regions with 
faculae.

The effect of a transiting planet is modelled as a circular dark spot. Therefore, the variations produced on the 
global CCF can be computed by

\begin{equation}
\Delta C^{tr}_{j}(v) = -\sum\limits_{k} C_{p}(v_{k}) \cdot H^{kj}_{p} \cdot p^{kj}_{tr},
\label{eqCtr} 
\end{equation}

As in Eqs.~\ref{eqtrans}, \ref{astroshX} and \ref{astroshY}, if the 
planet occults part of an active region, the corresponding fraction of 
the surface is computed within $p^{kj}_{tr}$ instead of the occulted 
$p^{kj}_{s}$ or $p^{kj}_{f}$.

Finally, the total CCF simulated for the observation $j$ is obtained by 
adding the contribution of the active regions (Eq.~\ref{eqCdeficit}) and 
the transiting planet (Eq.~\ref{eqCtr}) to the signature of the 
unspotted photosphere (Eq.~\ref{eqCim}):

\begin{equation}
C_{j}(v) = C^{im}(v) + \Delta C^{ar}_{j}(v) + \Delta C^{tr}_{j}(v).
\label{eqCtotal} 
\end{equation}

The function $C_{j}(v)$ is fitted with a Gaussian profile using a least 
square algorithm to obtain an accurate measurement of the 
peak, which is associated to the RV measurement for the 
time step $j$ of the simulation. The fit is made for the data contained 
in a window of width $\Delta v_{\rm CCF}$, specified in the input parameters, 
around the maximum of $C_{j}(v)$. A value of  20 to 40~km~s$^{-1}$ for 
$\Delta v_{\rm CCF}$ has been tested to be optimal for the simulations 
of Sun-like stars. This needs to be adjusted according to the stellar rotation and radius, ensuring that both wings and parts of the continuum are inside the window of the CCF. The FWHM of the Gaussian fit is also provided as an 
output of the simulations, as this is related to broadening mechanisms 
\citep{Gray92}. Finally, the bisector span is computed at each time step 
of the simulation from the difference between the mean of the 60 to $90\%$
interval of the bisector of the CCF in Eq.~\ref{eqCtotal} and the 10 to $40\%$
interval. This definition accounts for the inverse slope of the bisector
(BIS), as introduced by \cite{Queloz01}, and it is the most commonly used in
recent studies based on HARPS data \citep{Basturk11,Figueira13}.



Figure~\ref{simjitters} shows the spectroscopic measurements simulated for
a single active region on a Sun-like (left) and a K0 star (right), including a
comparison with simulations obtained with SOAP 2.0 for the same configurations.
Inspection of the RV panel (third from the top) shows that the introduction of
faculae ($Q=3$) notably increases the influence of stellar surface with inhibited
convective blueshift, thus producing a symmetric signal. The
$\sim$350~m~s$^{-1}$ mean difference between the quiet Sun and sunspot
bisectors is the main parameter producing the RV variations. The effects of
observed solar bisectors to account for the redshift in the active region
introduce asymmetries in the FWHM and the BIS signatures. 
While being of the same order, the differences with the SOAP
2.0 simulations in these measurements can be explained by a different approach
when modelling the center-to-limb variations of the bisectors, both for the
quiet photosphere and for the active regions. While SOAP 2.0 uses the same CCF
models across the full disk, in \texttt{StarSim} we use a more sophisticated 
and accurate approach of modelling the shape of the bisector
by interpolating CIFIST 3D models of different $\mu$ angles (see
Fig.~\ref{fig:bisectors}). Also the $\sim$350~m~s$^{-1}$ redshift of active
regions is scaled with $\mu$, accounting for the projection of the convective
motion. As discussed in Appendix B of \cite{Dumusque14a}, where a simple
limb-shift model of spectral lines is implemented, the approach of SOAP 2.0 can
produce spurious phase shifts in the spectroscopic measurements.

\begin{figure}[]
\begin{center}
\includegraphics[width=1.05\columnwidth,angle=0]{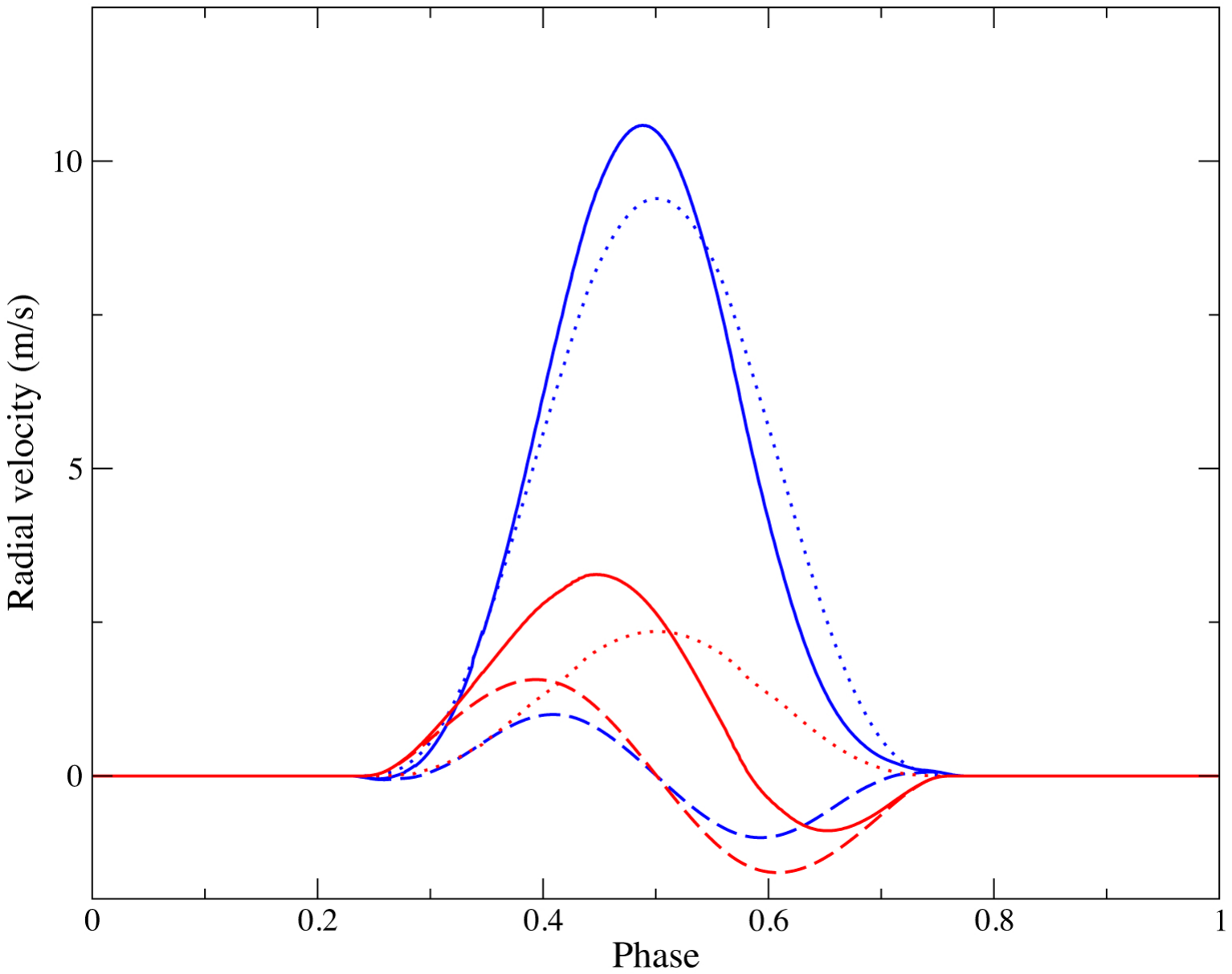}
\end{center}
\caption{RV shifts produced by a single spot at 
$\theta=90^{\circ}$ with a size of ${A}_{\rm Sn}=1.6\cdot10^{-3}$ 
stellar surfaces on a Sun-like star with $i_{*}=90^{\circ}$, both for a simulation without ($Q=0$, in red) and with faculae ($Q=3.0$, in blue). The result 
of simulating a rotating star ($P_{\rm rot}=25$ days) with no convection 
is plotted with a dashed line, showing the flux effect. The simulation of a convective star with 
no Doppler shifts is plotted with a dotted line. Both effects are 
included in the results plotted with a solid line. The HARPS 
passband is used to compute all the RVs.}
\label{spotrv}
\end{figure}

Figure~\ref{spotrv} shows the radial velocity signature of a single active
region on a Sun-like star. The effects of rotation (flux) and convection
are displayed separately, both for the case of a spot without faculae ($Q=0$) and
with surrounding faculae ($Q=3.0$). Notice that the flux effect dominates the
total RV signal for spots, while active regions with faculae
produce a convection dominated signature. While convection is blocked both for
spots and faculae, the temperature contrast is much larger in the first, so that the effect of rotation is more important. Those explanations and results can be
generalized to rapid rotators as shown by \cite{Dumusque14a}.

The \texttt{StarSim} tool also computes the level of Ca~{\sc ii} H\&K chromospheric emission
by means of the $S$ index \citep{Wilson78,Vaughan78,Duncan84}. The total
emission of the visible surface of the star is computed from the contribution
of all the surface elements, again accounting for the surface, projection and
limb darkening factors of each element (see Eq.~\ref{eqCimins2}). As before, to 
improve efficiency we first integrate the emission of a quiet star and then add
the contribution from the active regions and, if necessary, subtract the area
occulted by transiting planets (similarly to Eqs.~\ref{eqCim},
\ref{eqCdeficit}, \ref{eqCtr}, and \ref{eqCtotal} for the CCF). The $S$
emission from the active regions, as well as from the quiet photosphere
elements, were estimated from the measurements reported by \cite{Baliunas95}
and \cite{Hall07} for the quiet Sun ($S_{quiet}\sim0.164$), and for the active
Sun ($S_{act}\sim0.2$), assuming a $\sim0.3\%$ spot coverage and $Q=0$ and
$Q=3$ \citep{Balmaceda09,Foukal98,Chapman11}. Simulated time series
data of the $S$ index for a Sun-like star with a single
spot are displayed in the bottom panel of Fig.~\ref{simjitters}. Considering
the previous assumptions, the variability is in agreement with the measurements
by \cite{Baliunas95} comprising several complete solar cycles.

\section{Simulated spectral signature of active regions}
\label{specsigna}

As presented in Sect.~\ref{starsim}, \texttt{StarSim} allows to model the
photometric and spectroscopic variability of an active star with cool starspots
and hot faculae modulated by stellar rotation. The spectral signature of 
active regions is defined as the dependence of the flux variability or 
photometric jitter introduced by active regions with the equivalent 
wavelength of the observations. This can only be measured from time 
series spectroscopy or multi-band photometry of spotted stars and 
provides the key for the estimation of spots temperatures and sizes, 
which remain degenerated when observations in a single photometric band 
are available. \cite{Pont13} provide measurements of the temperature 
contrast of spots in HD~189733 (see also \citealt{Pont08,Sing11}) 
by computing the wavelength dependence of 
the amplitude of the flux increase caused by a spot occultation observed 
during a planetary transit, yielding $\Delta T_{\rm spot}\geq750$~K, 
which is consistent with a the mean projected filling factor of 1 - 3$\%$ 
computed from the variability in the MOST photometry
\citep{Croll07,Miller08,Boisse09}.

The methodology implemented in \texttt{StarSim} allows to model the 
spectral signature of spots and faculae from multi-band observations of 
activity effects. We may adopt the flux rms as an adequate measurement 
of the variability in a given light curve. Simulations for a test case
Sun-like star ($T_{\rm eff}=5777$~K, $\log g=4.5$ and solar abundances) 
with a rotation period of 25~days are performed in order to generate 
time series photometry for 12 different filters, ranging from the 
visible to the mid-IR (Johnson UBVRI, 2MASS JHK and Spitzer IRAC 
bands). The time series photometry covers 75 days with a cadence of 
1~day. The stellar surface is populated with a distribution of 3 to 8 
active regions composed of spots of an average size 
$\bar{A}_{Sn}=2\cdot10^{-3}$, in units of the total stellar surface. Two 
different scenarios are considered regarding the presence of faculae, 
assuming $Q=0.0$ and $Q=5.0$. The temperature contrasts are initially set to 
$\Delta T_{\rm spot}=350$~K and $\Delta T_{\rm fac}=30$~K. The active 
regions are located at a mean latitude of $\pm40^{\circ}$ and evolve 
with a typical lifetime of 40~days.

\begin{figure}[]
\begin{center}
\includegraphics[width=\columnwidth,angle=0]{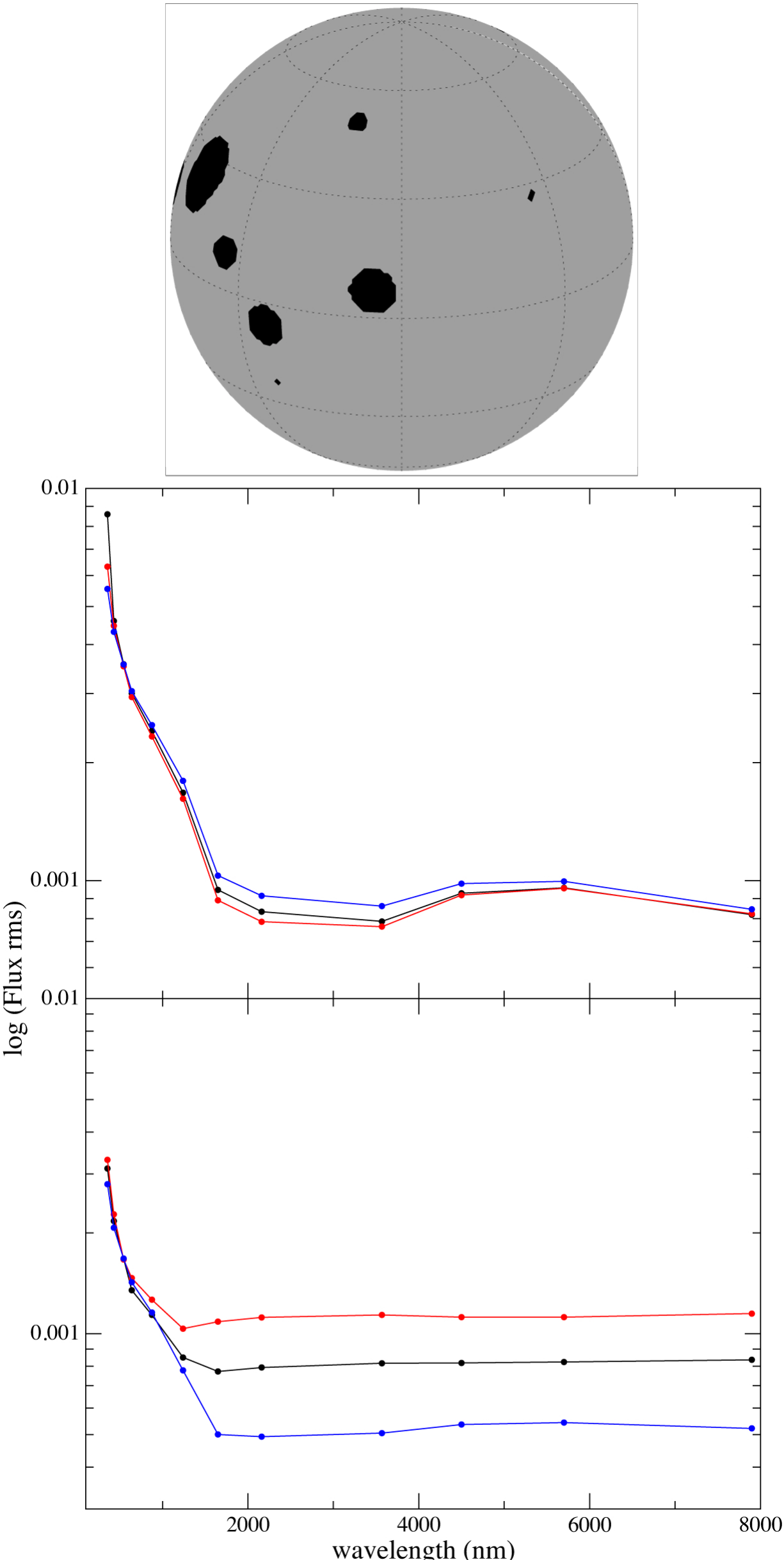}
\end{center}
\caption{Top panel: map of the initial distribution of active 
regions as used for the simulations in Sect.~\ref{specsigna} in the case 
of $Q=0.0$. Middle panel: the spectral signature (flux rms vs. 
wavelength) for three different configurations of active regions 
normalized to the same resulting rms in the Johnson V band: 
$\Delta T_{\rm spot}=310$~K and $\bar{A}_{Sn}=2\cdot10^{-3}$ (black 
line), $\Delta T_{\rm spot}^{+50\%}=230$~K and a $50\%$ increase in the area of 
the spots (red line), and $\Delta T_{\rm spot}^{-50\%}=730$~K and a $50\%$ 
decrease in the area of the spots (blue line). Bottom panel: same as 
the middle panel including facula regions with $Q=5.0$ and $\Delta 
T_{\rm fac}=30$~K, where the red line corresponds to $\Delta T_{\rm spot}^{+50\%}=230$~K and the blue line to $\Delta T_{\rm spot}^{-50\%}=550$~K.
}
\label{fig:specsigna}
\end{figure}

The resulting photometric signature for each configuration is computed in the
Johnson UBVRI, 2MASS JHK and Spitzer IRAC filter passbands, thus providing the
spectral signatures of the active regions computed from $\sim300$ to
$\sim8000$~nm. These are presented in Fig.~\ref{fig:specsigna}, together with a
snapshot of the stellar surface map at the initial time of the simulations without faculae. The same active region map is used for all the
simulations.  In a first step, a scenario with $Q=0.0$ was considered and
$\Delta T_{\rm spot}=350$~K and $\bar{A}_{Sn}=2\cdot10^{-3}$ were assumed. The
resulting spectral signature is plotted with a black line in the middle panel
of Fig.~\ref{fig:specsigna}. Then, we scaled the spot areas to $\pm50\%$ and we
searched for the $\Delta T_{\rm spot}$ values that preserve the flux rms in the
Johnson~V band. We obtained $\Delta T_{\rm spot}^{+50\%}=230$~K and $\Delta
T_{\rm spot}^{-50\%}=730$~K for the enlarged and reduced spots, respectively.
The resulting spectral signatures for these two configurations of temperature
contrasts and sizes of spots are also plotted in the middle panel of
Fig.~\ref{fig:specsigna} with a red line and a blue line, respectively. Whereas
the variability differs more than $2\cdot10^{-3}$ (relative flux units) in the
ultraviolet, the signature of the spots is barely distinguishable for the three
parameter configurations in the rest of the spectral range. The results change
significantly when faculae are included around the spots, as shown
in the bottom panel of Fig.~\ref{fig:specsigna}. We assume $Q=5.0$, which is
slightly lower than the mean solar value \citep{Chapman87,Lanza03}, and $\Delta
T_{\rm fac}=30$~K in all cases. Then, we recompute the spectral signature for
the same three configurations previously described. In this case, the spot
temperature contrasts that preserve the photometric variability in the
Johnson~V band are $\Delta T_{\rm spot}^{+50\%}=275$~K and $\Delta T_{\rm
spot}^{-50\%}=550$~K. The signature of active regions is dominated by the
emission of bright faculae in the near and mid infrared, especially for the
configurations with the lowest spot temperature contrasts, $\Delta T_{\rm
spot}=350$~K (black line) and $\Delta T_{\rm spot}=275$~K (red line), as we see
an increase in the rms after $\sim1500$~nm.

These results show that the determination of the basic spot parameters 
require spectrophotometric time series measurements at the $10^{-4}$ precision
level. On the other hand, the presence and amount of faculae can be more easily
measured in the red and near-IR bands, since there the photometric signature of the
simulated configurations differ by $\sim10^{-3}$.

\section{Modelling real data}
\label{photosim}

\subsection{Algorithm description}
\label{photosim_mod}

\texttt{StarSim} is intended to generate simulated signals from arbitrary distributions of stellar and activity parameters, as described in Sects.~\ref{sim_photo} and \ref{sim_rv}, with a better precision than other existing tools. The problem we are facing here is so called inverse problem, consisting in finding the causal factors of an observation. In this case, the aim is obtaining a dynamical spot configuration that fulfills the photometry time series. In the case of active stars, this could be a clue step for disentangling planetary signals, specially those existing in RV data. The methodology used in \texttt{StarSim} to simulate the stellar surface is also implemented to perform the inverse problem and obtain detailed
stellar surface maps that best fit the observed photometric signature
produced by $n-$ active regions. The observables are
described as
\begin{equation}
y_{i}(t_{i})\simeq f_{i}(s_{j},a_{k}) \hspace{5mm} (k=1,\ldots,n),
\label{eqalgor} 
\end{equation}
\noindent where $y_{i}$ is the measurement at time $t_{i}$,
$f_{i}$ is the model, $a_{k}$ are the stellar parameters, and $s_{j}$ is the configuration of active
regions. The generation of the functions $f_{i}$ is described in
Sect.~\ref{sim_photo} and Sect.~\ref{sim_rv} for the photometric and the spectroscopic observables, respectively.

In our case, \texttt{StarSim} performs a Monte Carlo Simulated Annealing
minimization (MC-SA). This is a generic probabilistic global optimizer that
allows to find a reasonably good approximation to the optimum in large
configuration-space problems. Although this method has been used mainly in
discrete spaces, it is also appropriate to tackle continous problems as the
case of continous parameters of the spot list. One of the significant
advantages of this optimization technique is the proven robustness against
local minima in noisy surfaces \citep{Kirkpatrick83}, which makes it an
interesting approach to optimize models involving a \emph{large} number of
parameters. The name of the algorithm --and its associated nomenclature--
comes from the term ``annealing'' in metallurgy, a technique involving a
scheduled cooling of metals to modify the size of crystals and reduce their
defects. 
  
The aim of the fitting process is to find a map of active regions that minimizes the $\chi^{2}$,
\begin{equation}
 \mathcal{S}^{*} = \arg \min_{s_j \in \boldsymbol{\mathcal{S}}} \mathrm{\chi^2}(\mbox{data}\,;\, s_j),
\label{eqargmin} 
\end{equation} 

\noindent where $\chi^{2}$ is evaluated between the model and the data,

\begin{equation}
	\chi^2 = \sum_{i=1}^{N}\frac{[y_i -  f_i]^2}{\sigma_i^2}
    \label{eqn:chi2_0}
\end{equation}

\noindent and $\sigma_i$ are the errors associated to de data.

The activity map $s_j$ is defined as a set of parameters containing the coordinates, initial time, lifetime and size of a preselected number of active regions, $n$.

$\mathcal{S}_j$ = \{ $t_{init}$, $t_{life}$, $\phi$ (colatitude), $\lambda$ (longitude), radii \},  $j=1,\ldots,n$

\noindent where $s_j$ is a specific configuration of active regions from all the possible sets, $\boldsymbol{\mathcal{S}}$, and $\mathcal{S}^*$ is the optimal configuration.

Further details on the MC-SA procedure are given in Appendix~\ref{annealing} together with the parameters as used in \texttt{StarSim}.

\subsection{Current status of StarSim}
\label{starsim_status}

The \texttt{StarSim} tool is currently capable to generate simulated photometric and spectroscopic signals as described in Sections~\ref{sim_photo} and \ref{sim_rv}. The \texttt{StarSim} code is currently written using a combination of Fortran
and C routines. The simulations ran on an Intel Core i7-3770 CPU at 3.40~GHz
produced a time series of 1000 phases in 60~seconds for photometric data (see
Sect.~\ref{sim_photo}) of a Sun-like star in the Johnson V band for a random
distribution of four spots. In the case of high-resolution spectroscopic data
(see Sect.~\ref{sim_rv}), the same simulation takes 210~seconds.

The algorithm to perform the inverse problem by means of a simulated annealing procedure (see Sect.~\ref{annealing}) is currently implemented only in the case of a single photometric time series. The inclusion of RV data, as well as multiple photometric passbands, would require a more sophisticated treatment of errors and is being under consideration for a new version of the tool. At the moment, simultaneous series data for active stars are being studied with \texttt{StarSim} by simulating a grid of RV solutions from the best fit of the photometry. An example including a discussion on the constrains provided by the analysis on the stellar and activity parameters is presented for the case of HD~189733 in next section.

\section{A case example: HD 189733}
\label{hd_ex}

\subsection{System parameters}
\label{hdparams}

The exoplanet host HD~189733 is a bright ($V = 7.67$~mag) K0V-type star known
to present significant rotation modulation in its light curve \citep{Winn07}
with a period of $\sim 11.9$~days and to be relatively active from the
measurement of the chromospheric activity indices \citep{Wright04,Moutou07}.
The existence of a hot Jupiter planet with a short period (2.2 days) was first
announced by \cite{Bouchy05}. Several studies have reported the presence of
activity effects on the RV measurements of up to
$\sim15$~m~s$^{-1}$ \citep{Lanza11c,Bouchy05} and have obtained information
about the stellar surface photosphere from the modelling of the flux variations
\citep{Croll07,Lanza11c}. The parameters for the HD~189733 system, as used in
all the simulations in this section, are presented in Table~\ref{tab:hdparams}.

\cite{Miller08} studied the transit timing variations (thereafter TTVs) and the
transit depth variations (thereafter TDVs) from 6 transit observations obtained
by MOST, which monitored HD~189733 for 21 days during July - August 2006. No
significant effects of spot crossing events were observed at any phase of the
transits, but no further discussion on the possible effects of non-occulted
spots was presented. When a transiting planet passes in front of a starspot,
the depth of the transit is shallower in a portion of the transit due to the
fact that the planet has crossed a dimmer region of the photosphere. In the
case of low signal-to-noise observations that cannot resolve the event, this
would affect the overall transit depth fit. Also, TTVs of $\sim$ 1 minute would
be induced if a spot crossing event occurred at the ingress or egress phases of
the transit. See \cite{Barros13} for an extensive discussion. On the
other hand, non-occulted spots present during the transit time introduce
variations of the stellar flux with chromatic dependence. Thereafter, they can
produce a bias in the transit depth measurement and modify the signature of the
atmosphere of the planet as studied by transmission spectroscopy. Since a spot
is modelled as a cooler region on the stellar surface, its signal will be
stronger in the blue, where the flux contrast increases, than in the red.

\begin{table}
\caption{Parameters for the system HD~189733}
\label{tab:hdparams}
\begin{tabular}{lll}
\hline\noalign{\smallskip}
  Parameter & Value & Reference \\
\noalign{\smallskip}\hline\noalign{\smallskip}
\hline
$T_{\rm eff}$ & 5050$\pm$50 K & M06 \\
$\log g$ & 4.53$\pm$0.14 & M06 \\
$\rm[Fe/H]$ & -0.03$\pm$0.04 & M06 \\
$\Delta T_{\rm spot}$ & $\sim$560 K & L11 \\
$Q$ & 0.0  & L11 \\
$P_{0}$ & 11.953$\pm$0.009 days & H08 \\
$i_{*}$ & $85.5^{\circ}\pm0.1^{\circ}$ & T09 \\
$A_{\rm Sn}$ & $5\cdot 10^{-4} - 3\cdot 10^{-3}$ & L11 \\
$P_{\rm planet}$ & 2.2185733$\pm$0.0000014 days & MR08 \\
$R_{\rm planet}/R_{\rm *}$ & 0.1572$\pm$0.0004 & P07 \\
$b$ & 0.671$\pm$0.008 & P07 \\
$\gamma$ & 1.4$\pm$1.1$^{\circ}$ & W06 \\
\noalign{\smallskip}\hline
\end{tabular}

References: M06 \citep{Melo06}; MR08 \citep{Miller08}; P07 \citep{Pont07}; L11
\citep{Lanza11c}; H08 \citep{Henry08}; T09 \citep{Triaud09}; W06
\citep{Winn06}.
\end{table}

\cite{Pont07} described some properties of the starspot groups in HD~189733
from the observation of two spot crossing events. The features observed in the
transit light curves show a wavelength dependence and can be explained by the
presence of cool regions with $\Delta T_{\rm spot}\sim$1000~K and a size of
$\sim 12000$~km to $\sim 80000$~km, which correspond to $A_{\rm Sn}$ of
$\sim2\cdot 10^{-5}$ to $\sim6\cdot 10^{-3}$ in units of the total stellar
surface.

\cite{Lanza11c} obtained a more detailed map of the active region distribution
by modelling MOST photometry covering 30.5~days from July 17 to August 17, 2007, and RV measurements from SOPHIE
obtained for the same time span \citep{Boisse09}. Their approach was based on reconstructing a
maximum entropy map with regularization. This allows to model the longitudinal
evolution of active regions over time, but the information on the latitudes is
lost particularly when the inclination of the stellar rotation axis is close to
$90^{\circ}$, as it is the case for HD~189733. So the method tends to assume that
the spots are located near the stellar equator, hence adopting the minimum
possible size for the spots to simulate the observed variations. The results of
\cite{Lanza11c} show that the data can be explained by the presence of 2 to 4
active regions yielding a photometric fit with an rms of $4.86\cdot 10^{-4}$
 in relative flux units and a resulting RV
curve that in rough agreement with the observed data (to
within a few m~s$^{-1}$). The individual spots cover 0.2 to 0.5\% of the
stellar disc and their lifetimes are comparable or longer than the duration of the
MOST observations (i.e., $\sim30$~days), while the rise and decay occur in 2-5
days. In this case, the spot contrast is computed assuming a spot effective
temperature of 4490~K (i.e., $\Delta T_{\rm spot}=560$~K). 
 A different spot contrast would imply a
change in the absolute spot coverage to explain the same photometric signature
\citep{Pont07}, but it would not have a significant effect in a single band
photometric time series. 
\cite{Fares10} measured the equatorial and polar
rotation periods obtaining $11.94\pm0.16$~days and $16.53\pm2.43$~days,
respectively, thus giving a relative amplitude of
$\Delta\Omega/\Omega=0.39\pm0.18$, which is very similar to the Sun (so we
assume $k_{\rm rot}=1$). Then, the migration rates of spots measured by
\cite{Lanza11c} indicate that most of the spots are located at latitudes
ranging from $40^{\circ}$ to $80^{\circ}$.

\subsection{Activity model}
\label{hdphoto}

We analyze the MOST photometry dataset from 2007 \citep{Lanza11c}.
After removing the transits, the series contains 414 measurements with a mean
error of $1.15\times10^{-5}$ in relative flux units. The RV data
from the SOPHIE spectrograph were obtained from July 13 to August 23, 2007,
providing a data set of 33 RV measurements. The combination of these
simultaneous data sets represents a great opportunity to test the methodology
of \texttt{StarSim} to obtain a surface model of activity for HD~189733.

\begin{figure*}[]
\centering
    \resizebox{\hsize}{!}{\includegraphics{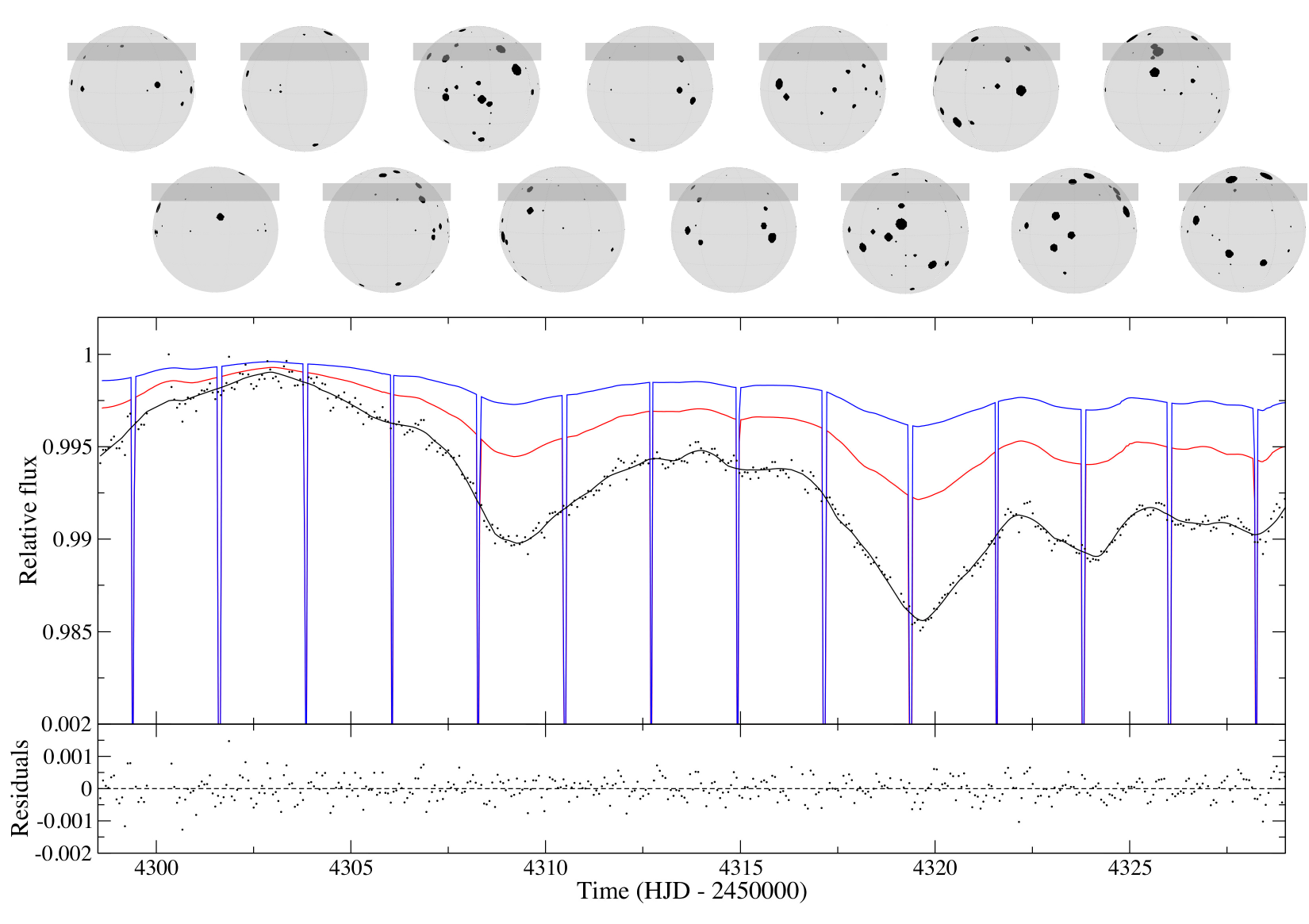}}
    \caption{Synthetic light curves generated for HD~189733 with the
methodology described in Sect.~\ref{sim_photo}, assuming the parameters
described in Sect.~\ref{hdparams} and the modelled activity map, for the MOST
(black line), 2 MASS J (red line) and IRAC 2 (blue line) passbands. The flux is
in relative units with respect to the maximum along the light curve and shows
the 14 analysed transit events. Observed data from MOST is plotted with black
dots. The projected maps of the stellar surface at the mid-transit times are
plotted in the upper part, including a dark grey band that indicates the region
of the star occulted by the planet.} 
     \label{hd_rms1}
\end{figure*}

The \texttt{StarSim} program with the minimization methodology presented in
Sect.~\ref{photosim_mod} was used in order to obtain a surface map for
HD~189733. In a first approach, the MOST photometric dataset was modeled, no
faculae were considered (see \citealt{Lanza11c}), and the spot sizes were
assumed to evolve with the mean solar growth/decay rate (i.e. $a_{\rm
rd}\simeq1^{\circ}$, \citealt{MartinezPillet93,Petrovay99}). A map of 90 active
regions with a grid resolution of $1^{\circ}\times1^{\circ}$ was fitted for
their position, time of appearance and lifetime, while the stellar parameters
were fixed to the ones specified in Table~\ref{tab:hdparams}. Despite
considering a relatively high number of active regions, the minimization
algorithm tends to concentrate the spots in groups, as expected. The number was
adjusted to reduce the rms of the fit while preventing isolated small spots to
appear in the modelled surface. Three maps containing 90 spots were randomly generated and used as initial conditions for three different fitting runs and the simulated annealing procedure was permormed as described in Sect.~\ref{annealing} in order to determine the convergence. The final rms of the residuals is
$3.8\cdot10^{-4}$ in relative flux units for all the solutions, which only present visible differences for the latitudes and sizes of some of the spot groups. The light curve model for the first run is shown in
Fig.~\ref{hd_rms1}.

\begin{figure}[]
\begin{center}
\includegraphics[width=1.02\columnwidth,angle=0]{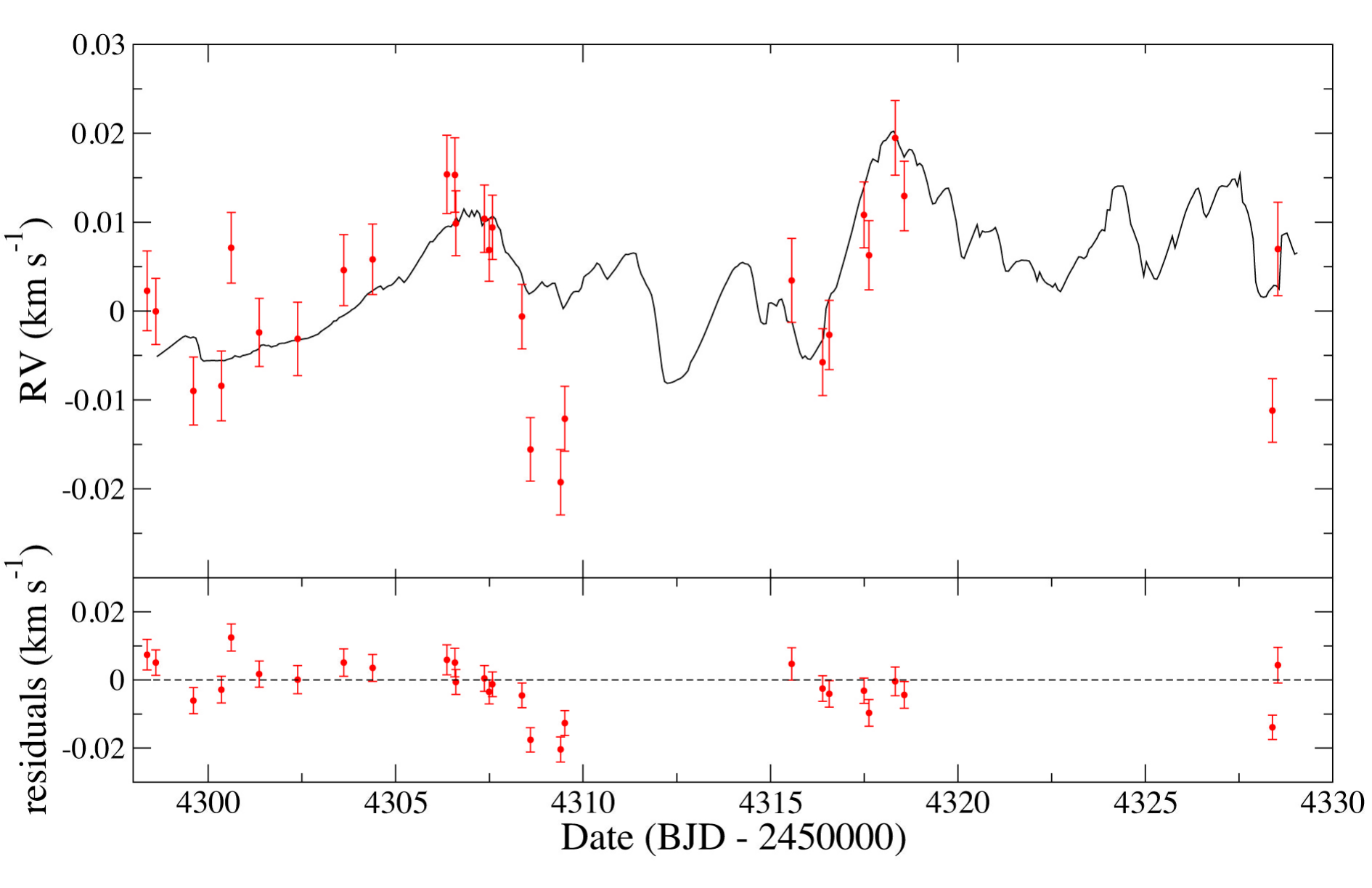}
\end{center}
    \caption{Upper panel: RV model curve generated with \texttt{StarSim} for HD~189733 (black solid line) compared to the SOPHIE data (red dots). Bottom panel: residuals between the data and the model.
    } 
\label{hd_rv90}
\end{figure}

The photometric fit that we obtain is of better quality than those published to
date and our model is able to describe with precision many of the features of
the photometry. The distribution of active regions derived from the
photometric fit and the parameters defined in Table~\ref{tab:hdparams} were
adopted in order to synthesize the RV curve for the same time span and for the
instrumental passband of SOPHIE. The result is plotted in Fig.~\ref{hd_rv90}
together with the SOPHIE data. The model and the observed RV variations are
generally in good agreement except for the regions around $\sim$HJD 2454309 and
$\sim$HJD 2454327, where probably a higher evolution rate for the spots is
needed in order to reproduce the rapid modulations. However, we could not
reproduce them when exploring different initial parameters for the active
regions, so they are either caused by instrumental effects, or such variations
are not explained by spots and faculae. The analyses presented by
\cite{Lanza11c} and \cite{Aigrain12} using photometric spot models were also
unable to reproduce the rapid RV variations. When excluding the three
measurements between HJD 2454307 and 2454310, the rms of the residuals of our
model is 5.45~m~s$^{-1}$, which improves the results from the previous studies (5.55~m~s$^{-1}$ from \citealt{Lanza11c} and 6.6~m~s$^{-1}$ from \citealt{Aigrain12})
and is close to the typical instrumental uncertainties of SOPHIE
($\sim$5~m~s$^{-1}$).
 
A further step forward in the analysis would be to include the SOPHIE RVs
in the inverse problem and to perform a simultaneous fit. However, several
tests indicated that the number of data points and the precision of the
velocity data are too low for the fitting algorithm to work properly.
Alternatively, we used the RVs to investigate the effect of several parameters
that are usually adopted {\em ad hoc} or from independent criteria because of
the lack of sensitivity of the light curve. The parameters are the zero point
of the photometric flux (i.e., immaculate flux level), the stellar equatorial
rotation period ($P_{\rm rot}$), the facula-to-spot area ratio ($Q$) and the
temperature contrast of the active regions.

\begin{figure*}
\centering
    \resizebox{0.95\hsize}{!}{\includegraphics{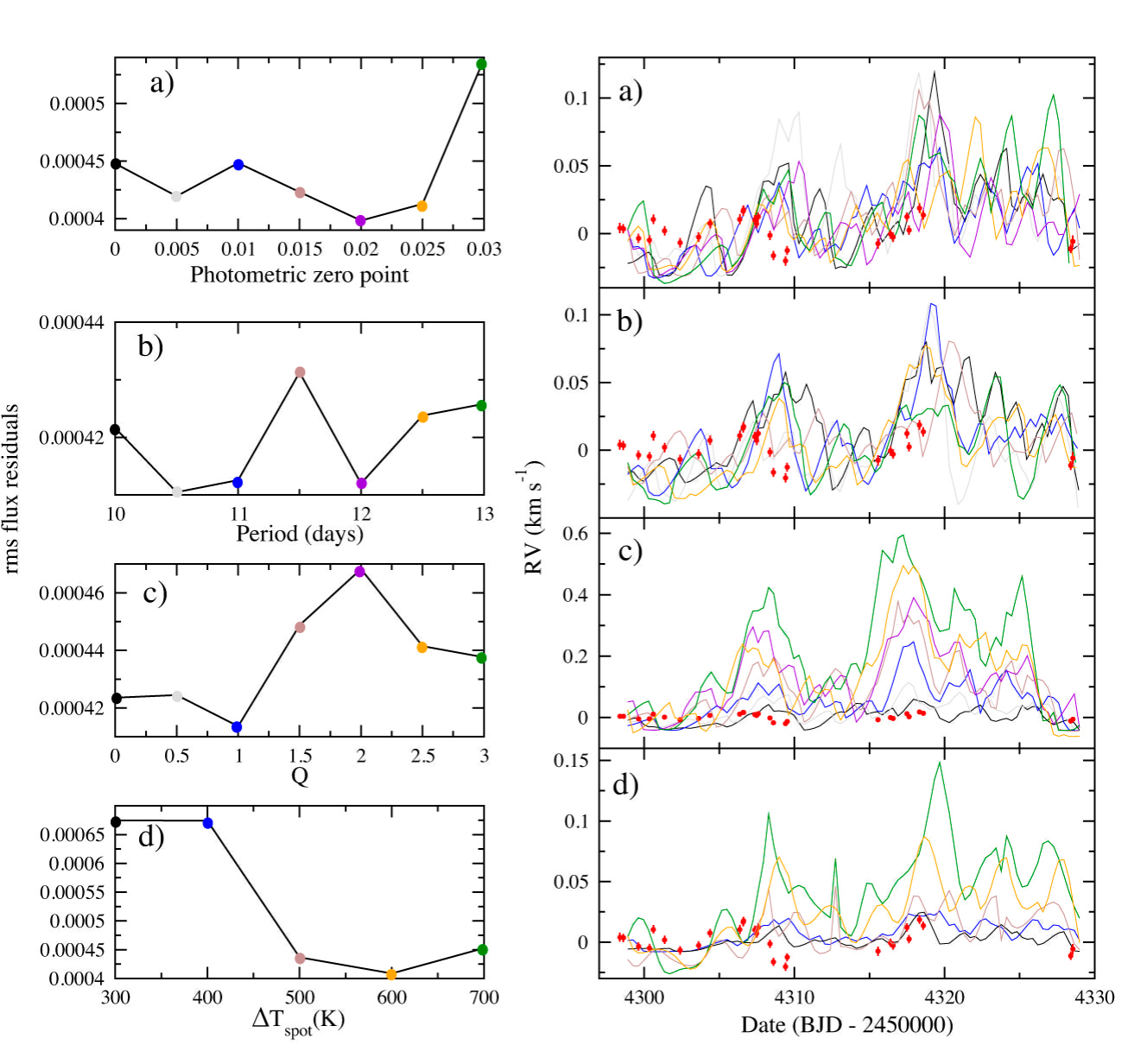}}
    \caption{Left panels: rms of the residuals from the fit to the MOST light
curve with a model of 30 active regions vs. four stellar parameters. Right
panels: simulated RV curves resulting from the models compared to the SOPHIE
data (red dots). The four pairs of panels show the dependence of the rms of the light curve fits and the RV
signature on a) the zero point of the photometry, b) the stellar rotation
period, $P_{\rm rot}$, c) the facula-to-spot area ratio, $Q$, and d) the
temperature contrast of the spots, $\Delta T_{\rm spot}$. The colour code of the lines in the right panels correspond to the dots in the left panels.} 
     \label{hd_rms2}
\end{figure*}

The {\texttt StarSim} code was used to fit the photometry with the same
assumptions as described before and by exploring a grid of possible values for
the four parameters in sequential and independent fashion. In this case, a
model of 30 active regions was used for simplicity. The results are shown in
Fig.~\ref{hd_rms2}. The panels on the left show the rms of the photometric fit
as a function of the parameter value, while the panels on the right display the
simulated RV curves compared with the observed data. As expected, all
investigated parameters have very little influence on the quality of the
photometric fit except for the spot temperature contrast, which shows a broad,
flat minimum around $\Delta T_{\rm spots}\sim 600$~K, in agreement with
\cite{Lanza11c}. Regarding the impact on the RVs, it is observed that the
variations in the photometric zero point and the rotational period of the star
induce phase differences in the simulated RV curves. Figure~\ref{hd_rms2} also
shows that the value of $Q$ has a very strong influence on the RV amplitude.
The photometric fit, although not very significantly, favours values of $Q$
close to 0, and this is also in agreement with the modest observed RV
variation. The amplitude of the RV signal is also sensitive to the spot
temperature contrast. The results of the simulations shown in the right
panel d) indicate that $\Delta T_{\rm spots}\sim 400-500$~K are in close
agreement with the SOPHIE data, while higher values show increasing
deviations.

We see notable differences in the RV simulations adopting different parameters,
thus indicating that the degeneracies present could be partially broken if RV
data are considered in the inverse problem. Note that the initial
conditions of the distribution of active regions (positions and sizes) are
randombly chosen and could also have an impact on the simulated RV curves shown
in the right panels of Fig.~\ref{hd_rms2}. This, however, is not a concern when
the full inverse problem is considered, since different random realizations can
be used to explore the entire parameter space. In this case the modelling
would allow for a full description of the stellar surface, including active
region parameters, and thus a complete characterization of the stellar activity
properties. The application of this procedure to a sample of stars would help
to understand some key aspects and therefore provide means to correct RV
measurements from the unwanted effects of stellar activity in a self-consistent
way.

\subsection{Activity effects in transit photometry}
\label{hd_photo}

Synthetic light curves for HD~189733 including planetary transits were
generated using the methodology presented in Sect.~\ref{sim_photo} for the 12 filters used in Sect.~\ref{specsigna} for the same time span as the MOST photometric data modelled in Sect.~\ref{hdphoto}. This covers three complete
rotation periods of the star (i.e., 14 transits of the exoplanet HD~189733b).
The surface map from the photometric model obtained in Sect.~\ref{hdphoto},
consisting of 90 spots, was considered for all the simulations. The resulting light curve has a maximum amplitude of
$\sim0.012$ mag in the Johnson~V filter. The light curves for the 2~MASS~J and the IRAC~2 passbands are plotted in Fig.~\ref{hd_rms1}.

For each passband, data segments of $2\cdot T_{14}$ (i.e., twice the total
duration of the transit) around the transit mid point were selected for
analysis. The out-of-transit baselines at either side of the transit were used
to fit a parabolic trend that was subsequently subtracted from the in-transit
observations. The data points affected by spot crossing events were removed and
not considered for analysis. Then, a \cite{Nelson72} model as implemented by \cite{Popper81} was fitted to each
individual transit using the JKTEBOP code \citep{Southworth04} to measure the transit depth and
the mid transit time. In a first step, precise quadratic limb darkening
coefficients were obtained for each passband by fitting a model to a transit
light curve generated for an immaculate photosphere. With these limb darkening
coefficients, the full series of transits was analysed by fitting the planet
radius and the zero point of the ephemeris. The rest of the parameters were
fixed to the ones shown in Table~\ref{tab:hdparams}.

A number of spot crossing events have been detected in HD~189733b transit
observations \citep{Miller08,Sing11,Pont13}. The observed occurrence of these
events combined with the assumption that the planet is crossing a typical zone
of the stellar surface in terms of spot coverage, leads to an estimate of
$1-2\%$ of projected spot filling factor \citep{Sing11}, which is in agreement
with the results by \cite{Lanza11c} and the spot maps used in this work. 
In the case of our synthetic data, as the active region map is poorly
constrained due to the low signal-to-noise ratio of the MOST observations (see
\citealt{Lanza11c}), and especially the latitudes of the spots are degenerated, we do
not expect all the events from the MOST data to be accurately reproduced by the model. For this reason, the data affected by spot crossing events were removed before fitting the transit models in order to study and quantify the chromatic effects of non-occulted spots.

\begin{figure}[]
\begin{center}
\includegraphics[width=1.02\columnwidth,angle=0]{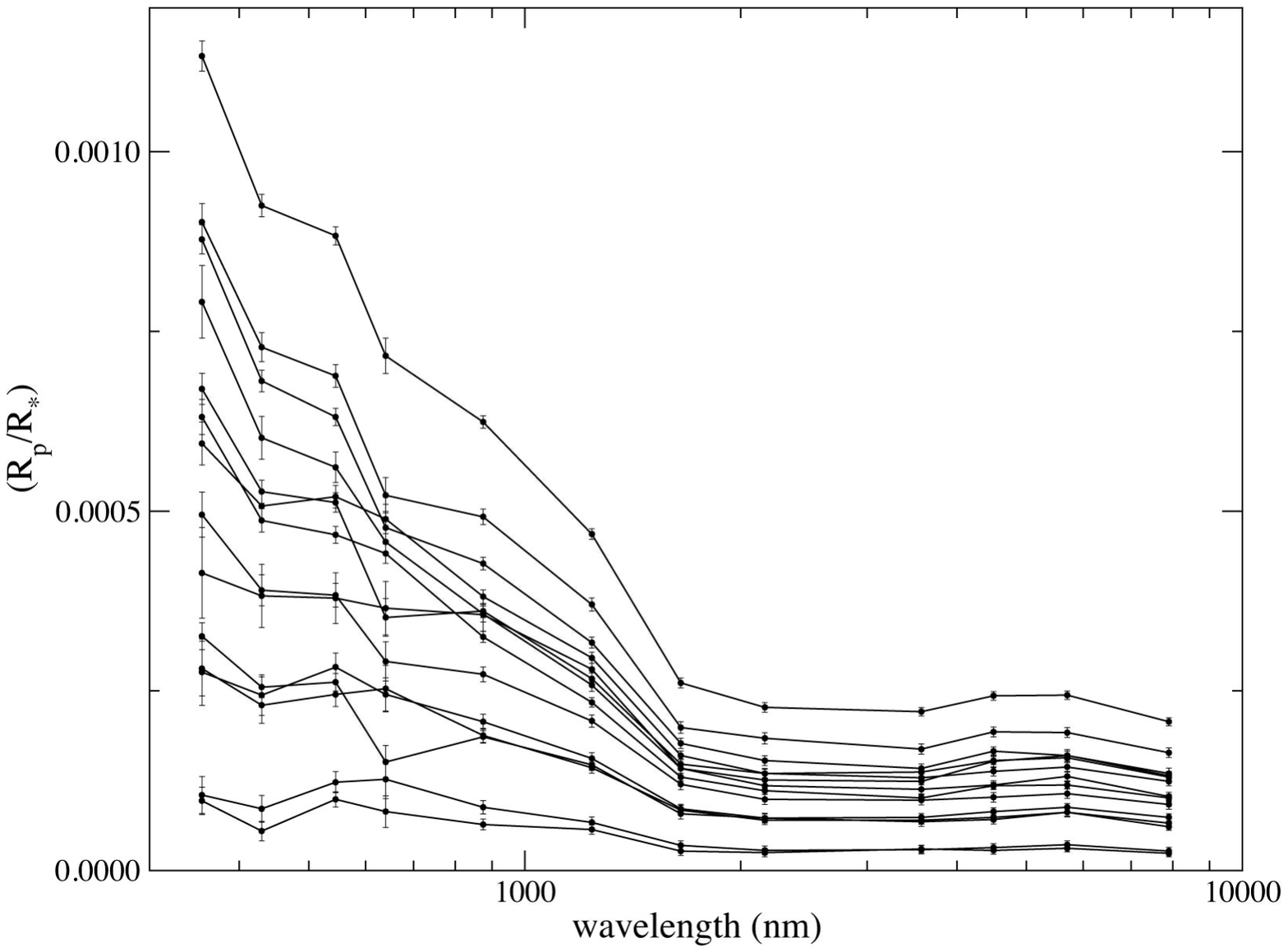}
\end{center}
    \caption{Variations on the transit depth $R_{p}/R_{*}$ vs. the central
wavelength of 12 different filters produced by several configurations of spots
on the 14 synthetic transits of HD~189733b in the simulated light curves,
relative to the transit depth in an unperturbed transit. Data affected by spot
crossing events have been removed.
    } 
\label{hd_all}
\end{figure}

The results of the analysis of synthetic transit data are displayed in
Fig.\ref{hd_all}, showing the chromatic dependence of the planet radius
determination for each transit event. All transits present some level of influence
from non-occulted spots. The measurements for $R_{p}/R_{*}$ are normalized
at the values obtained for the transit light curve simulated on an unperturbed
photosphere.

\begin{figure}[]
\begin{center}
\includegraphics[width=1.02\columnwidth,angle=0]{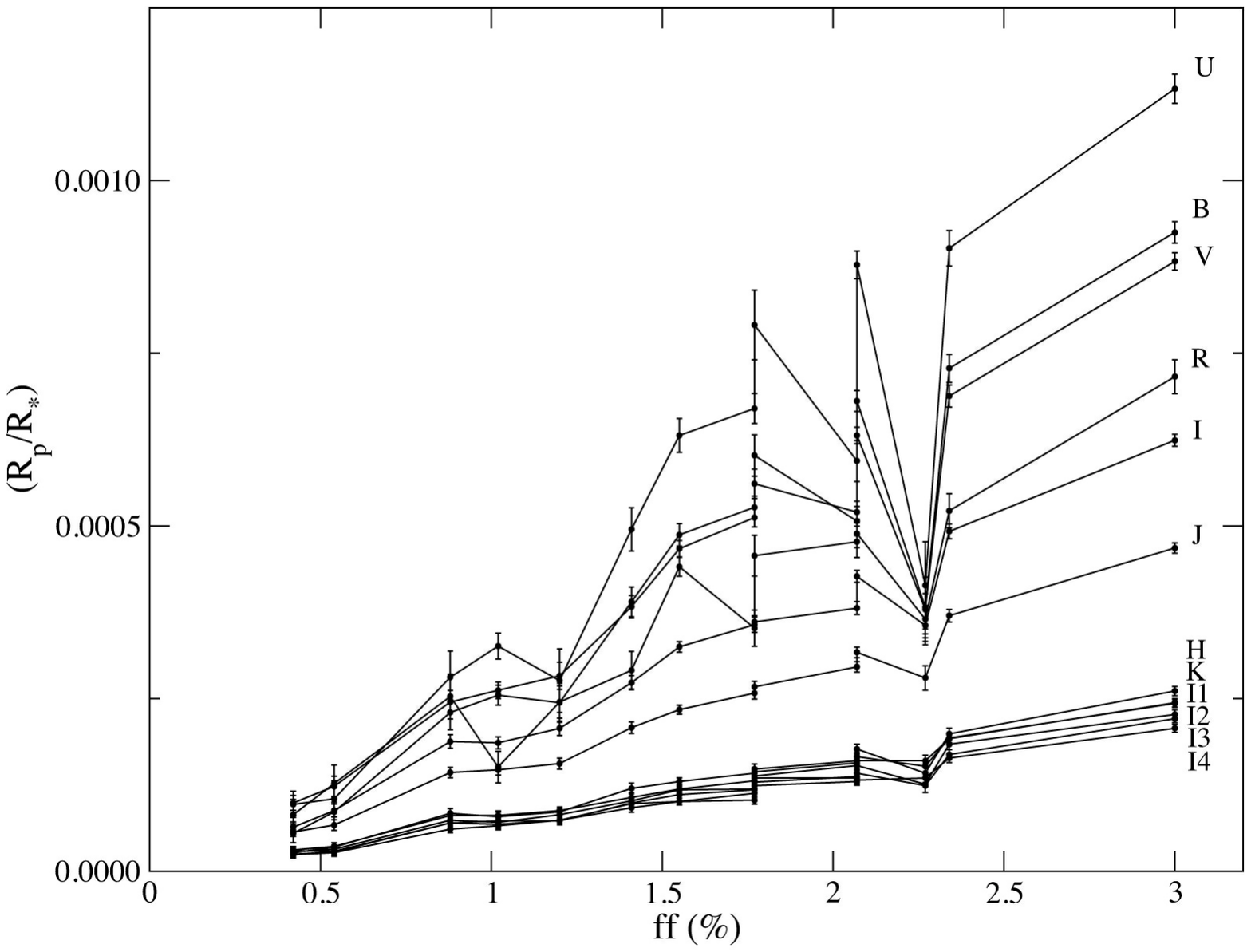}
\end{center}
    \caption{Variations on the transit depth $R_{p}/R_{*}$ vs. the  projected
filling factor of spots at the mid transit times for the 12 simulated
passbands.
} 
\label{hd_ff}
\end{figure}

As the
presence of spots in the stellar surface changes the overall transit depth
measurement (i.e. the duration of the transit is much shorter than the typical changes in the projected filling factor),
there are no significant differences on the measured times of mid-transit. Spots not occulted by the planet introduce a monotonic regular behavior from
the UV to mid-infrared wavelengths. 
Therefore, the spectral signature of the spots clearly dominates the effects
introduced on the transit depth, as seen in Fig.~\ref{hd_all}. The results for
the transit depth measurements are also displayed vs. the projected filling
factor of spots (measured at the mid-transit times) in Fig.~\ref{hd_ff} for the
12 analysed filter passbands. The dispersion in this plot is due to the gaps introduced in the data when removing spot crossing events.
For all transit simulations, the effect produced by spots is much stronger in
the blue than in the mid-infrared, as a consequence of the spectral
signature of the spots. The shift on the transit depth measurement
introduced by non-occulted spots is scaled with the instantaneous filling
factor, ranging from $\Delta\left(R_{p}/R_{*}\right)\sim0.0001$ for a nearly
unspotted surface (3rd transit, at HJD$\sim$2454301.8 in Fig.~\ref{hd_rms1}) to
$\sim0.001$ for a projected filling factor of $\sim3\%$ (10th transit, near
HJD$\sim$2454319.0). This is an important information to take into account when
correcting the transit depths for activity, as observations of HD~189733 show
that there is significant variability in the filling factor over time.
Note that in order to obtain the spot filling factors affecting our simulations (shown in
Fig.~\ref{hd_ff}) we assumed that the maximum of the MOST photometry as the unit of measure for the flux in order to normalize the light curve \citep{Lanza11c}.
Although the evolution of the spot map we used accurately reproduces the
variations, a zero point in the filling factor of spots could be present. This is not still clear from the results shown in Fig.~\ref{hd_rms2}, as discussed in Sect.~\ref{hdphoto}.
However, the statistics of spot crossing events during HST observations and the
characteristics of the transmission spectrum in the visible \citep{Sing11},
indicate that the projected spots filling factor is not much higher than
1-2~$\%$.


\section{Discussion and conclusions}
\label{sim:conc}

We present a methodology, in the form of the \texttt{StarSim} tool, to simulate the
photosphere of rotating spotted stars. This allows us to generate synthetic 
photometric and spectroscopic time series data from a set of given 
parameters and a library of stellar atmosphere models. The main purposes 
of this methodology are the characterization of the effects produced by 
activity in time series observations, especially multi-band studies of 
exoplanetary transits, and the design of strategies for the minimization 
of activity effects, both when searching and characterizing 
exoplanets. 
Several aspects should be taken into account to discuss the 
reliability of the results and their interpretation.

First, we assumed LTE stellar models for the emission of the quiet 
photosphere and also for the spot and facula regions. This is a 
limitation imposed by the availability of stellar model libraries with 
high resolution spectra. The synthetic spectrum shown in 
Fig.~\ref{comp_sun} correctly fits the observed spectrum of a spot and 
provides a temperature contrast in agreement with most of the values in 
literature (see Sect~\ref{starsimmodels}). However, certain lines 
are not well reproduced, which suggests that the spot spectrum is different 
from a cooler stellar photosphere. In the case of faculae, no high 
resolution observed spectra are currently available, and the most recent 
NLTE models for the Sun are the ones presented by \cite{Fontenla09}. 
The differences in the spectrum of active regions mainly affect the 
existence of certain lines, but have no significant impact on the flux 
when considering a broad range of the spectrum. Therefore, the currently 
used models are suitable for the characterization of activity effects in 
the observations for most of the photometric instrumentation. Some more 
considerations on the influence of the models for the different 
photospheric features will be discussed in a future paper addressing 
strategies for the correction of RV measurements in active 
stars. Such RV measurements are usually obtained by using the information
content of the lines in the whole observed spectral range. On the other hand,
optimized measurements can be made by excluding the spectral lines or ranges
that are most affected by activity. 

Second, given the requirements of this work, the maps of active regions
currently used in our simulations are limited to a model of circular spots with
a surrounding facula in the shape of a corona, instead of a more complex
structure with groups of small spots. Such simple configurations of active
regions have proven to adequately reproduce photometric and spectroscopic
signature in active stars \citep{Eker94,Lanza03,Dumusque14a,Dumusque14b}.
However, \texttt{StarSim} also allows modelling spots of any shape by including an
unlimited number of spots of a small size on the stellar surface. Our approach
is very similar to that of \cite{Lanza03} and subsequent papers that use this
type of spot maps to model the surface of active stars. In such case, the
regularization of the maximum entropy algorithm provides a smooth distribution
of active regions instead of a more irregular surface map populated with a lot
of small spots. 

Third, several assumptions based on solar data are made in our
simulations. This has to be kept in mind when using \texttt{StarSim} for stars 
other than Sun-like. Indeed, the limb brightening law implemented for facula
regions is a model based on solar observations. The same applies to the rest of
the parameters of faculae, such as the temperature contrast. The evolution
rates and lifetimes of active regions have only been measured for a few stars
from the various techniques that allow modelling of the stellar
surface \citep{Hussain02,Strassmeier94a,Strassmeier94b,Petrovay97,Petrovay99}.
While the previous assumptions affect some parameters that can be adjusted when
more information from observations is available, our approach to model activity
effects on convection is more intimately linked to
solar observations in all cases. The center-to-limb convection profile of the
quiet photosphere is implemented from solar CIFIST 3D models.
Also, solar high-resolution observations are used to
obtain accurate models for the bisector of active regions. Our approach to
model convection was tested to perform at the $\sim$1~m~s$^{-1}$ precision in
some comparisons with Sun-like stars. Finally,
the chromospheric emission data produced by our model is based on solar
observations of the $S$ index.

An example of the capabilities of the methodology for the 
characterization of activity effects is presented for the case of a 
Sun-like star with a typical distribution of active regions. The spectral signature (flux rms vs. wavelength) 
of the active regions is studied in visible and IR passbands for 
several configurations of the parameters of the stellar surface. 
Different configurations of the temperature contrast and the typical area of the spots are found to 
show a very similar signature when no faculae are introduced. In the 
presence of faculae ($Q=5.0$, $\Delta T_{\rm fac}=30$~K), the wavelength 
dependence is inverted in the red and near-IR wavelengths when the 
temperature contrast of the spots is $<300$~K, thus being the 
contribution of faculae dominant. In all cases, the dependence is strong 
in the visible and near-IR, while no signature is observed in the 
mid-IR. Real observations of the spectral signature for active 
stars in visible and near-IR filters may provide new estimations 
of the facula-to-spot area ratio and the parameters of the 
spots.

The effects of non-occulted spots on multiband observations of planetary 
transits were studied by considering the case of the active Sun-like star HD~189733
with a transiting hot Jupiter planet. A Monte Carlo Simulated Annealing minimization algorithm was implemented in \texttt{StarSim} and stellar surface maps were obtained by modelling the MOST photometric data for HD~189733. Several stellar parameters which are usually assumed were explored for degeneracies by using simultaneous RV data from SOPHIE. We show that the RV data includes some additional information that can break these degeneracies, especially for the physical parameters of the active regions and the stellar surface map.

Considering the modelled surface map for HD~189733, photometric light curves were 
synthetised for 14 transits in 12 different passbands from the optical to 
the mid-IR, covering a variety of spot surface coverages from $\sim0.4\%$ to
$\sim3.0\%$, and the deviations produced on the planet radius measurements were
studied. While in the first case the transit depth 
measurement remains unperturbed up to the $10^{-4}$ precision in 
$R_{p}/R_{*}$ for any filter passband, a strong effect is introduced 
when the star is moderately spotted. Moreover, the difference is wavelength 
dependent, being $\sim10^{-3}$ in the blue, and only $\sim5\cdot10^{-4}$ in the 
mid-IR.  Therefore, biases of the order of $10^{-3}$
can be introduced in $R_{p}/R_{*}$ measurements for wavelengths in the optical range if a constant spot filling factor is assumed for the non-occulted
spot correction \citep{Sing11,Pont13,McCullough14} instead of considering the 
appropriate instantaneous spot map. Time series photometry simultaneous to 
the transit observation campaign would be needed to reproduce the evolution of
the filling factor of spots and adequately scale the correction for each
transit event. In the near future, space telescopes dedicated to exoplanetary sciences will be able to achieve precisions of $\sim10^{-4}$ on the atmosphere measurements of super-Earths \citep{Tessenyi12}, so it is essential to adequately model the spectral signature of activity for such types of observations.

We conclude that the approach presented in this paper to generate 
synthetic time series data accurately accounts for all the physical 
processes affecting spectroscopic and photometric variations produced by 
activity, which currently represents the main obstacle for exoplanet 
characterization and searches. In fact, our approach can help to model 
and correct for the effects of activity jitter on photometric and RV observations of extrasolar planets, especially if the system 
parameters are known and the surface map of the stellar host is simultaneously studied. In comparison with already existing tools to model time series data of active stars \citep{Oshagh13,Boisse12,Aigrain12,Dumusque14a}, \texttt{StarSim} accurately accounts for the line bisector distortions caused by convective blueshift and convection inhibition in active regions, based on recent 3D atmosphere models. Most of the cited tools assume a simple Gaussian model for the spectral lines. In the case of \texttt{StarSim}, we consider atmosphere models in order to compute the synthetic CCFs and the flux contribution of the visible surface elements of the star. This allows to preserve all the spectral information of the variability produced by activity effects. Also, our simulations allow studying the effects of the 
amount of faculae as well as of spot temperature contrasts and sizes 
on multi-band simultaneous photometry. The methodology is 
currently being implemented 
to model fit multi-band photometric and spectroscopic RV observations 
simultaneously, thus obtaining detailed maps for the characterization of 
stellar photospheres. Several applications of the methodology to real 
observations and to the design of optimized strategies for the search and 
characterization of exoplanets will be presented in a series of papers 
in the near future. \texttt{StarSim} will also help to 
properly account for stellar activity in the detection and 
characterization of atmospheres in small planets.

\begin{acknowledgements}  
E. H. and I. R. acknowledge support from the Spanish Ministry of Economy and
Competitiveness (MINECO) and the Fondo Europeo de Desarrollo Regional (FEDER)
through grants ESP2013-48391-C4-1-R and ESP2014-57495-C2-2-R.  C. J.
acknowledges the support by the MINECO - FEDER through grants
AYA2009-14648-C02-01, AYA2010-12176-E, AYA2012-39551-C02-01 and CONSOLIDER
CSD2007-00050. E.~H. was supported by a JAE Pre-Doc grant (CSIC).
\end{acknowledgements}

\bibliographystyle{aa} 
\bibliography{biblio.bib}

\begin{appendix} \section{Simulated annealing procedure}
\label{annealing}

We show here a simple outline of simulated annealing optimization method
implemented in \texttt{StarSim}.

\begin{enumerate}
\item Draw a random configuration (a list of random spots). 
\begin{itemize}
\item Set initial inverse temperature $\beta = \beta_0$, where $T=1/\beta$ is defined as the cooling temperature of the algorithm.
\item Let $k=0$ (number of temperature steps).   
\item Let $n_{\beta} = 0 $ (number of steps for each temperature)
\end{itemize}
\item Make the current configuration $\phi_j$ evolve to a perturbed
configuration $\phi_j^p$ by \emph{slightly} modifying one randomly selected
parameter of one of the spots. 

\item If $\chi^{2}(\phi_j^p)$ $<$ $\chi^{2}(\phi_j)$ then\\
\hspace*{4mm} accept $\phi_j^p$ \\ else \\ \hspace*{4mm} accept
$\phi_j^p$ with probability $e^{-\beta \Delta \chi^{2}}$ \\ $n_{\beta}
\leftarrow n_{\beta}+1$. If $n_{\beta} \leq n_{\beta}^{MAX}$ go to 2. 

\item Increase the inverse temperature $\beta \leftarrow \beta + \delta \beta$ according to a \emph{cooling schedule}.
\item $k \leftarrow k+1$. If $k < k_{max}$ go to 2, else go to 6.
\item End.

\end{enumerate}

The great advantage of MC-SA is the ability of avoiding local extrema given an
adequate \emph{cooling schedule}, which is the way the temperature is
decreased. Hence, a favourable configuration after a perturbation is always
accepted while a disfavourable perturbation can be accepted or not according to
an exponential probability, making it possible to \emph{hillclimb} to finally
reach the global optimum. There is no consistent rule to choose an
optimal annealing schedule as it strongly depends on the specific problem to
optimize. However, a fairly efficient scheme may be determined by trial and
error \citep{Kirkpatrick83} to obtain a robust minimizer, and fast
convergence was found for a Kirkpatrick cooling schedule,
\begin{equation}
 \beta_k = \beta_0 \, \alpha^{-k},
\label{kirkpatrick}
\end{equation} 
\noindent in comparison to a linear scheme, both widely used since they were
first introduced \citep{Nourani98}. Several trials were carried out to adjust
the parameters ($T_{0}$, $\alpha$, $n_{\beta}$ and $k$) until proper behavior
was reached. Table \ref{tab:MCSAparams} summarizes the values of all MC-SA
parameters as configured in \texttt{StarSim}.

\begin{table}
\caption{Parameters for the MC-SA algorithm as used in \texttt{StarSim}}
\label{tab:MCSAparams}
\centering
\begin{tabular}{ll}
\hline\noalign{\smallskip}
  Parameter & Value \\
\noalign{\smallskip}\hline\noalign{\smallskip}
\hline
$T_{0}$ & $1000$ \\
$\alpha$ & $0.25$\\
$n_{\beta}$ & $1000-10000$ \\
$k$ & $20$ \\

\noalign{\smallskip}\hline
\end{tabular}
\end{table}

\end{appendix}

\end{document}